%% file: main.tex
\documentclass[aps,pra]{revtex4-1}

\usepackage{amsmath} 
\usepackage{amssymb}
\usepackage[dvipsnames]{xcolor} 
\usepackage{amsfonts}
\usepackage{placeins} 

\usepackage{enumerate} 

\usepackage[ruled]{algorithm2e} 

\usepackage{graphicx} 
\usepackage[final]{hyperref} 
\hypersetup{
	colorlinks=true,       
	linkcolor=blue,        
	citecolor=blue,        
	filecolor=magenta,     
	urlcolor=blue         
}

\input{macros.tex}

\begin{document}

\title{Ensemble-filtered vortex modeling of strongly disturbed aerodynamic flows}
\author{Mathieu Le Provost and Jeff D. Eldredge}
\email{jdeldre@g.ucla.edu}
\affiliation{Mechanical \& Aerospace Engineering Department, University of California, Los Angeles, Los Angeles, CA, 90095, USA}

\begin{abstract}
The task of dynamic flow estimation is to construct an instantaneous approximation of an evolving flow---and particularly, its response to disturbances---using measurements from available sensors. Building from previous work by Darakananda et al.~(Phys Rev Fluids 2018), we further develop an ensemble Kalman filter (EnKF) framework for aerodynamic flows based on an ensemble of randomly-perturbed inviscid vortex models of flow about an infinitely-thin plate. In the forecast step, vortex elements in each ensemble member are advected by the flow and new elements are released from each edge of the plate; the elements are modestly aggregated to maintain an efficient representation. The vortex elements and leading edge constraint are corrected in the analysis step by assimilating the surface pressure differences across the plate measured from the truth system. We show that the overall framework can be physically interpreted as a series of adjustments to the position and shape of an elliptical region of uncertainty associated with each vortex element. In this work, we compare the previously-used stochastic EnKF with the ensemble transform Kalman filter (ETKF), which uses a deterministic analysis step. We examine the response of the flat plate at $20^\circ$ in two perturbed flows, with truth data obtained from high-fidelity Navier--Stokes simulation at Reynolds number 500. In the first case, we apply a sequence of large-amplitude pulses near the leading edge of the plate to mimic flow actuation. In the second, we place the plate in a vortex street wake behind a cylinder. In both cases, we show that the vortex-based framework accurately estimates the pressure distribution and normal force, with no {\em a priori} knowledge of the perturbations or their structure. We show that, in each case, the ETKF is consistently more robust than the stochastic EnKF and is qualitatively better at representing the coherent structures of the true flow. Finally, we examine the mapping from measurements to state update in the analysis step through singular value decomposition of the Kalman gain.

\begin{description}
\item[Keywords]inviscid vortex model, ensemble Kalman filter, disturbed separated flow, flow estimation
\end{description}
\end{abstract}

\maketitle


\section{\label{sec:intro}Introduction}


The objective of aerodynamic flow estimation is to construct an approximation of the time-varying flow about a lifting surface from the available sensors. This task has a variety of potential uses, principally within the context of an overall closed-loop control strategy, in which the estimation of the evolving flow field enriches the information available to the controller. For example, we may seek to exploit transient flow mechanisms, such as the high lift achievable in dynamic stall \cite{dickinson1993unsteady, birch2001spanwise, milano2005uncovering,taira2009three,eldredge2019leading}, or to manage the flow in the presence of unknown atmospheric disturbances (gusts) \cite{kerstens2011closed,perrotta2017unsteady}. In all of these sitations, the flow is subjected to transient disturbances---from flow actuators, wing maneuvers, or gusts---and in many instances, the disturbance induces the flow to separate or significantly affects an already separated flow.

The flow estimation problem can be posed in a variety of settings, loosely categorized by the degree to which the estimator is built on a physics-based model. At one extreme, the estimator incorporates little to no physics and a supervised machine learning architecture is trained to map sensor data to some description of the flow field \cite{fukami2020assessment,hou2019machine}. In contrast, if the estimator relies on a predictive physics-based model, then its job is to use the sensor data to correct that prediction for unrepresented effects. In particular, though the physics model may be outfitted with the geometry and the basic flow state, the sensor data is the only source of information on disturbances. This prediction and measurement update process is the generic task of data assimilation, the focus of this paper.

All estimation strategies seek the state vector $\state_k$, which contains sufficient information to specify the system's state at any discrete time step $t_k$. In flow estimation, $\state_k$ not only contains a finite-dimensional representation of the flow field, but also any boundary conditions or model parameters that are uncertain. This state vector's evolution is predicted ({\em forecast}) with a dynamical operator $\dyn_k$, based on a time discretization of some underlying physics model,
\begin{equation}
    \label{eqn:dyn_basic}
    \state_k = \dyn_k(\state_{k-1}).
\end{equation}
The dynamical model is paired with an observation operator $\obs_k$,
\begin{equation}
    \label{eqn:obs_basic}
    \meas_k = \obs \left(\state_k \right),
\end{equation}
which returns a vector of predicted observations, $\meas_k$. The data assimilation framework provides a means of folding new observations from the true system into the model through adjustments of the state vector. Generically, this adjustment ({\em analysis}) is carried out through an optimization: determine the adjustment that minimizes the difference between new sensor measurements $\BB{y}^\star$ from the truth system and the model-predicted measurements from (\ref{eqn:obs_basic}). Though this optimal adjustment can be derived deterministically, it is important to remember that we anticipate uncertainty in the estimation task due to random noise in both the dynamics and the measurements. Thus, in stochastic estimation, such as in the various forms of Kalman filter, the adjusted state estimate is the one that minimizes its uncertainty after new measurements have been assimilated \cite{kalman1960new,asch2016data}. The degree to which we favor the dynamical prediction or the measurements is determined by the relative levels of noise we assume for each.

Under the assumption that the noises are drawn from Gaussian distributions, the system's uncertainty is completely characterized by the covariance matrix. Both the Kalman filter and its variant for non-linear systems, the extended Kalman filter, require that this covariance matrix be stored and propagated along with the mean state itself. For high-dimensional systems such as in fluid dynamics, this storage and propagation of the covariance matrix is computationally intractable. This challenge was addressed by Evensen with the ensemble Kalman filter (EnKF) \cite{evensen1994sequential}, a Monte Carlo approach to the two-step filter algorithm. In the EnKF, the system's mean and covariance are approximated by an ensemble of states randomly sampled from the probability distribution. In the forecast, each state is advanced by the dynamical model; in the analysis step, each ensemble member is updated with the new observation to minimize the posterior covariance. Thus, for state dimension $n$ and ensemble size $M < n$, one need only store $M n$ state components rather than a $n\times n$ matrix.

The dynamical operator in a flow estimation framework can take various forms, including high-fidelity Navier--Stokes simulation. Indeed, this has been the basis for recent work by da Silva and Colonius \cite{da2018ensemble,da2020flow}. In \cite{da2018ensemble}, they advanced the state vector for flow past an airfoil with an operator derived from an immersed boundary projection method \cite{taira2007immersed}. The state vector contained grid velocity data as well as the value of the freestream velocity and its time derivative. The truth system in this case was generated by the same simulations, but with the time-varying freestream velocity prescribed. However, the EnKF-based estimator only used sparse surface pressure measurements from this truth system, from which it successfully estimated the model's instantaneous freestream velocity. 

Since an ostensible goal for flow estimation is real-time control, dynamical models obtained from CFD will generally not be fast enough to update the state. In their subsequent work in \cite{da2020flow}, da Silva and Colonius utilized a dynamical model for the airfoil flow consisting of cheaper coarse-grid simulations. Their state vector was then augmented with a bias error, which they successfully estimated along with the coarsened grid state from surface observations of the higher-fidelity truth data. Though this coarse-grid approach is certainly faster, it is unclear whether the approach can be extended to very small-dimensional state vectors before aliasing errors would overwhelm the solution of the governing flow (Navier--Stokes or Euler) equations. Data-based modal decompositions, such as proper orthogonal decomposition (POD), are commonly used to reduce the dimension of grid-based solutions \cite{taira2017modal}, and a Galerkin model based on such a decomposition might provide a suitable approach for computationally-efficient estimation. However, a modal decomposition cannot efficiently capture the influence of large-amplitude flow disturbances if they are missing from the basis modes.

We seek an alternative low-dimensional representation of the flow, with a dynamical model that is computationally efficient but remains sufficiently rich to describe important physical phenomena. Aerodynamics has long relied on inviscid vortex models for representing and interpreting unsteady flow phenomena \cite{vonkarman38:1j,brown1955slender,sarpkaya1975inviscid,cottet2000vortex,eldredgebook}. Vortex models have a particular advantage in a flow estimation framework, because their physical fidelity and numerical accuracy are distinct. For example, even a model consisting of a small number of point vortices is still an exact solution (to within time marching error) of the two-dimensional inviscid equations of motion, including the non-linear convection. Thus, a vortex model of any size might serve as a low-dimensional representation of a more complex flow furthermore, it can readily account for the influence of disturbances, e.g., through Biot--Savart interactions.

Viscosity is obviously explicitly absent from an inviscid vortex model. However, among the roles of viscosity in moderate to large Reynolds number external flows, only the generation of vorticity at the surface of a body is truly essential to the flow's large-scale dynamics. Classically, this vorticity generation was restricted to vortex shedding from the trailing edge, by regularizing to flow at the edge with the Kutta condition \cite{vonkarman38:1j,eldredgebook}. However, in recent years, a number of inviscid vortex models have been developed for predicting aerodynamic flows with leading-edge separation, at a wide range of angles of attack \cite{wang2013low,xiamohseni17,ramesh2014discrete,darakananda2019versatile}. Ramesh et al.~\citep{ramesh2014discrete} introduced the concept of the leading-edge suction parameter (\lesp{}), a nondimensional measure of the amount of suction (integrated pressure) at the nose of an airfoil. In their criterion \cite{ramesh2014discrete}, which can be interpreted as a generalization of the Kutta condition \cite{eldredgebook}, vortex elements are only shed from the leading edge if \lesp{} exceeds a critical value, \lespc{}; if \lespc{} is set to zero, then the Kutta condition is enforced at that edge. The circulation of the new vortex element is proportional to the amount by which \lesp{} exceeds \lespc{}. Thus, to complete a basic inviscid vortex model of a separated flow past an airfoil, one need only specify the value of \lespc{}. Ramesh et al.~\citep{ramesh2014discrete} hypothesized that this critical value is constant and a function only of the airfoil shape; they tuned its value in a thin airfoil model to match the observed separation in a high-fidelity Navier--Stokes simulation of the airfoil. This hypothesis has been confirmed to hold for a wide variety of unsteady flows generated by an airfoil undergoing unsteady maneuvers \cite{ramesh2018leading}. However, for flows subjected to disturbances such as gusts, there is no {\em a priori} reason to believe that the hypothesis still holds, and, without further knowledge, its value should be assumed time-varying and uncertain. Furthermore, because \lespc{} is a measure of a critical behavior in the pressure, its value is likely to be observable from surface pressure measurements.

Thus, an inviscid vortex model outfitted with a leading-edge shedding condition based on \lespc{} is a strong candidate for the forecast step in aerodynamic flow estimation. A caveat of the model is that its computational complexity is $O(q^2)$, where $q$ is the number of elements. With elements shed from the edges at every time step, the model can easily lose its computational advantage over CFD-based approaches. To restrain the growth of this dimensionality, vortex elements can be aggregated with various conservation constraints, e.g., based on moments of the vortex system \cite{spalart1988vortex}, the rate of change of linear impulse \cite{darakananda2019versatile}, or the velocity induced at the leading edge \cite{sureshbabu2019model}. Aggregation addresses another issue that arises when a vortex model is applied to a flow past an airfoil at moderate angle of attack. Such flows exhibit extensive interactions between shed vorticity and the plate. Singular vortex elements thus remain close to the plate and, without the influence of viscosity, create strong and non-physical pressure disturbances that interfere with the model's predictive utility. Aggregation tends to move vortex elements away from the plate, thereby mitigating this behavior \cite{darakananda2019versatile}.

To exploit these potential benefits of a vortex-based assimilation procedure, Darakananda et al.~\cite{darakananda2018data} developed a data-assimilated vortex model in which the state vector consisted of the positions and strengths of vortex elements and the value of \lespc{} for a flat-plate airfoil. The vortices were aggregated by the same approach as in \cite{darakananda2019versatile}, which ensured that the number of elements remained lower than 60 in all cases considered. The vortex model was expressed in an EnKF framework, with the observation vector composed from the jump in surface pressures across the flat plate. By assimilating measurements from a high-fidelity simulation at Reynolds number 500, the estimation framework was able to successfully reproduce the force history on the plate. 

The current work extends the estimation framework of Darakananda et al.~\cite{darakananda2018data} and is motivated by two observations from that work. In one of the examples in \cite{darakananda2018data}, the flat plate was subjected to vortical gusts generated upstream. Remarkably, the estimation framework performed well even with no representation for the gust itself in the state vector. In other words, the dynamics of interaction between the gust and the separated flow was entirely accounted for through adjustments made in the analysis steps to vorticity shed from the plate. The estimator needed no knowledge of the gust's structure, a great advantage since gusts are, by definition, unknown. In this work, we probe this more deeply by focusing exclusively on flows subjected to incident gusts. We purposely apply gusts that are stronger and in faster sequences than in \cite{darakananda2018data}, so that their individual influences on the flow cannot easily be decorrelated. Even if we knew each gust's parametric form, it would be very difficult to unambiguously disentangle their parameters. In this paper, we explore whether this parameter estimation task can be avoided.

The second observation from \cite{darakananda2018data} was that the estimator's performance on any given flow varied substantially from one realization to the next, often exhibiting non-physical spikes in the pressure and force at random instants. This lack of consistent behavior is clearly undesirable for practical estimation applications. We attempt to address this by utilizing a variant of the EnKF called the ensemble transform Kalman filter (ETKF). As we review later, the ensemble in the original EnKF (called the stochastic EnKF, or \senkf{}) does not reproduce the exact posterior covariance relation of the Kalman filter, causing spurious noise in the resulting estimate. The ETKF, developed by Bishop et al.~\cite{bishop2001adaptive}, fixes this with an analysis step that exactly reproduces the expected covariance.

The rest of this paper is organized as follows. Section \ref{sec:vortex} reviews the aggregated vortex model and presents a state-space formulation of the model. Section \ref{sec:enkf} introduces the ensemble Kalman filter theory, including both the original \senkf{} and the \etkf{}. The results of the data-assimilated aggregated vortex model, using both forms of EnKF, are presented on example problems in \ref{sec:results}. Concluding remarks follow in \ref{sec:conclusion}. For the sake of completeness, pseudo-codes for the two EnKF methods are provided in the Appendix.

\section{\label{sec:vortex}The aggregated vortex model}

This section provides a summary of the aggregated vortex model, which constitutes the dynamical part of our flow estimator. We focus in this paper on the response of a two-dimensional infinitely-thin plate of chord length $c$, translating impulsively from rest at a velocity $U$ in a fluid of density $\rho$. Throughout this study, positions will be reported in chord lengths and time will be measured in convective units $t^\star = tU/c$. 

The flow is modeled by a collection of $q$ regularized point vortices, called blobs, identical to the model used by Darakananda et al.~\cite{darakananda2018data}. The vorticity field is given, as usual for a regularized vortex particle method, by
\begin{equation}
    \omega(\BB{y},t) = \sum_{j=1}^{q} \Gamma_q \zeta_\varepsilon(\BB{y} - \BB{y}_q),
\end{equation}
where $\zeta_\varepsilon$ is a blob kernel with radius $\varepsilon$. The purpose of a blob is to de-singularize the Biot--Savart interactions with other nearby elements. Its effect on elements farther than $\varepsilon$ is indistinguishable from that of a point vortex. The blob kernel in this work takes the commonly-used algebraic form $\zeta_\varepsilon(\BB{y}) = (\varepsilon^2/\pi)(|\BB{y}|^2+\varepsilon^2)^{-2}$. We refer readers to \cite{eldredgebook} for a review of modeling strategies of two dimensional inviscid flows.

The Kutta condition is applied at the trailing edge. The shedding criterion based on \lespc{}, developed by Ramesh et al. \citep{ramesh2014discrete}, is applied at the leading edge. The \lespc{} is included with the vortex elements' positions and strengths as part of the system state vector, $\state$. Following Ramesh et al.~\cite{ramesh2014discrete}, the \lespc{} is forecast to remain constant. Ramesh et al.~\cite{ramesh2018leading} and Darakananda et al.~\cite{darakananda2018data} have shown that \lespc{} has a strong authority over the vortex dynamics near the leading edge. The true vortex dynamics are encoded in the pressure measurements of the plate, from which we can distill an estimate of the \lespc{} \citep{hou2019machine, darakananda2018data}. In other words, though \lespc{} is predicted to remain constant, the assimilated pressure measurements will tend to cause it to vary, thereby triggering the release of weaker or stronger vorticity.

We use the same vortex element aggregation scheme developed by Darakananda et al.~\cite{darakananda2018data}. By aggregating vortex elements at every time step, the overall population remains modest (smaller than 60 in all cases explored) and the aggregated elements stay relatively farther from the plate than without such treatment, dramatically reducing the occurrence of spurious pressure disturbances.

In our flow estimator, the state variable $\state_k$  contains the positions and circulations of $q$ blobs and the critical leading-edge suction parameter \lespc{},
\begin{equation}
    \state_{k} = \left[\begin{array}{llllllll}
x_{k}^{1} & y_{k}^{1} & \Gamma_{k}^{1} & \ldots & x_{k}^{q} & y_{k}^{q} & \Gamma_{k}^{q} & \mbox{LESPc}_{k}
\end{array}\right]^\top.
\end{equation}
Thus, the state vector dimension is $n = 3q + 1$. The dynamical model (\ref{eqn:dyn_basic}) applies the following operations:
\begin{enumerate}[(a)]
    \item Enforce the no-flow-through condition on the plate,
    \item Introduce new vortices according to the Kutta condition at the trailing edge and the current estimate of \lespc{} at the leading edge, 
    \item Advect vortices and plate,
    \item Aggregate vortex elements, and set strengths of aggregated blobs to zero.
\end{enumerate}
More details on $\dyn_k$ can be found in \citep{darakananda2018data}. The observation operator $\obs_k$ in (\ref{eqn:obs_basic}) uses the unsteady Bernoulli equation to predict the pressure jump coefficients---$\Delta C_p = 2(p^+ - p^-)/\rho U^2$ where $+$ and $-$ denote the upper and lower side of the plate---at $d$ locations on the plate \cite{darakananda2017vortex}. In this work, pressure measurements are calculated (and also provided by the truth system) at the following $d=50$ sampled Chebyshev points:
\begin{equation}
    \frac{c}{2}\cos \left(\frac{10i \pi}{512 + 1}\right) \quad \mbox{for }i=1,\ldots,d. 
\end{equation}


\begin{figure}[t]
\centering
    \vskip 0.5cm
    \includegraphics[width = 0.95\linewidth]{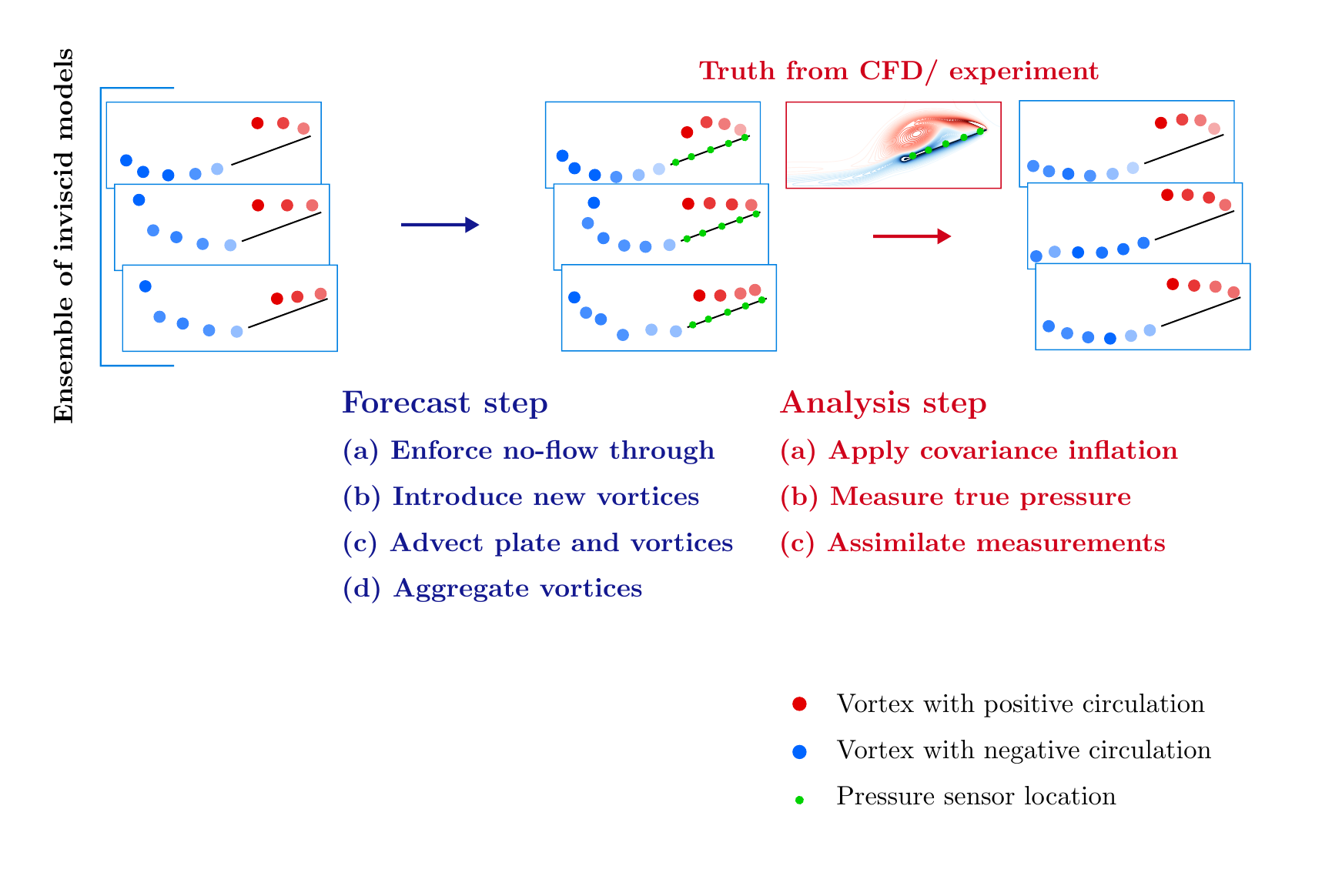}
    \caption{One time step of the data-assimilated aggregated vortex model}
    \label{fig:algo_vortex}
\end{figure}

\section{\label{sec:enkf}Ensemble Kalman filtering methodology}

In this section we describe the filtering methodology that underpins our data-assimilated vortex model. We start by reviewing the basic filtering problem and the purpose of addressing this problem with an ensemble approach. We then present two flavors of ensemble Kalman filter: the \senkf{} as well as a deterministic variant called the ensemble transform Kalman filter (\etkf{}) (Bishop et al.~\cite{bishop2001adaptive}). Both will evaluated in the vortex model applications that follow in Section~\ref{sec:results}. Then, we will review the important technique of covariance inflation to prevent filter divergence. Finally, we will interpret the data-assimilated vortex model on physical grounds.

\subsection{\label{sub:filtering}A review of the filtering problem}
In this section, we present a basic outline of the general filtering problem and its connection with Bayesian inference; details on this can be found in several references, including \cite{asch2016data}. Though the variants of the ensemble Kalman filter can be explained without this background, it is useful to discuss it in order to justify the use of the ensemble Kalman filter in a filtering problem and to identify some of its limitations. In this section and those that follow,
we use the following font conventions. Serif fonts refer to random variables, e.g. $\BB{\mathsf{Q}}$ on $\real^n$, or $\mathsf{Q}$ on $\real$. Lowercase roman fonts refer to realization of random variables, e.g. $\BB{q}$ on $\real^n$, or $q$ on $\real$. $\pi_{\BB{\mathsf{Q}}}$ denotes the probability distribution for the random variable $\BB{\mathsf{Q}}$, and $\BB{q} \sim \pi_{\BB{\mathsf{Q}}}$ means that $\BB{q}$ is a realization of $\BB{\mathsf{Q}}$. $\pi_{\BB{\mathsf{Q}} \given \BB{\mathsf{R}} =  \BB{r}}$, or $\pi_{\BB{\mathsf{Q}} \given \BB{\mathsf{R}}}(\cdot \given \BB{r})$, is the probability distribution of the random variable $\BB{\mathsf{Q}}$ knowing that random variable $\BB{\mathsf{R}}$ takes the value $\BB{r}$. It is called the \textit{conditional probability} of $\BB{\mathsf{Q}}$ given $\BB{r}$.

For the sake of generality, we frame our discussion on a generic discrete nonlinear state-space model (\ref{eqn:dyn_basic}), but we will connect the reader throughout to the present case of a low-order vortex model. Formally, the model is described by the pair of random variables $(\Statex_k, \Meas_k)$ for $k\geq 1$, where $\Statex_k$ is the state variable of a Markov process in $\real^n$ and $\Meas_k \in \real^d$ is an observation of the state $\Statex_k$, assumed to be conditionally independent of the state (detailed later in this section). In our case, the state variable is composed of the \lespc{} and the positions and strengths of the entire set of point vortices. The dynamics of the state $\Statex_k$ is described by the probability distribution for the initial condition $\pi_{\Statex_0}$ and a propagation equation, also called the dynamical equation:
\begin{equation}
\label{eqn:dynamic}
    \Statex_{k} = \dyn_k(\Statex_{k-1}) + \noisedyn_k,
\end{equation}
where $\dyn_k :\real^n \xrightarrow{} \real^n$ is the forward operator and $\noisedyn_k$ is an additive process noise. In the present case, $\dyn_k$ is the time-discretized vortex model. Propagating the dynamical equation for the state variable $\Statex_{k-1}$ is equivalent to sampling from the transition distribution $\pi_{\Statex_{k} \given \Statex_{k-1}}(\cdot \given \state_{k-1})$ where $\state_{k-1}$ is one realization of the random variable $\Statex_{k-1}$.

In general, the state $\Statex_{k}$ is only indirectly observed in a noisy and nonlinear fashion through $\Meas_k$:
\begin{equation}
\label{eqn:obs}
    \Meas_k = \obs_k(\Statex_k) + \noiseobs_k,
\end{equation}
where $\obs_k :\real^n \xrightarrow{} \real^d$ is the \textit{observation operator} and $\noiseobs_k$ is an additive \textit{measurement noise}. In our case, $\obs_k$ applies the unsteady Bernoulli equation to the state vector to obtain a vector of pressure differences, at discrete locations, between the upper and lower surfaces of the plate. The gradient of the observation operator $\obs_k$ evaluated at the current state $\state_k$ is called the \textit{tangent linear} of the observation operator, denoted $\Obs_k$, i.e., $\Obs_k = \nabla \obs_k$. It is important to note that we do not restrict ourselves in this section to Gaussian distributions for the initial condition, process noise, or measurement noise. Furthermore, the forward and observation operators can be time-dependent.

In the filtering problem, we leverage the realizations of the observation process $(\meas)_{1:k}$ to infer the realization of the state variable $\state$ at time step $k$. Our main objective is to estimate the posterior distribution of this state---providing us with complete information about the state and its uncertainty---given all of the observations made thus far, $\pi_{\Statex_k \given \Meas_{1:k}}(\state \given \meas_{1:k}) = \pi_{\Statex_k \given \Meas_1, \Meas_2,  \cdots ,\Meas_k}(\state \given \meas_1, \meas_2, \cdots ,\meas_k)$. Let us consider the estimation of this posterior distribution from the most recent observation, $\meas_k$, which can be computed from Bayes' theorem:
\begin{equation}
\label{eqn:bayes}
   \pi_{\Statex_k \given \Meas_{1:k}}(\state \given \meas_{1:k}) = \frac{\pi_{\Meas_{k} \given \Statex_k}(\meas_{k} \given \state)
   \pi_{\Statex_k \given \Meas_{1:k-1}}(\state \given \meas_{1:k-1})}{ \pi_{\Meas_{k}}(\meas_{k})},
\end{equation}
where $\pi_{\Meas_{k} \given \Statex_k}(\meas_{k} \given \state)$ is the likelihood distribution, i.e., the probability of the observation $\meas_k$ if we knew the state $\state$; $\pi_{\Statex_k \given \Meas_{1:k-1}}(\state \given \meas_{1:k-1})$ is the prior distribution, our estimate of the state before knowledge of the new observation $\meas_{k}$; and $\pi_{\Meas_{k}}(\meas_{k})$ is the distribution of the observation. Note that, in the likelihood, we have assumed that the observation errors are independent in time, so it is only conditioned on the current state and not on past observations. Since $\pi_{\Meas_{k}}(\meas_{k})$ can be thought of as a normalizing constant, we can rewrite (\ref{eqn:bayes}) as:
\begin{equation}
\label{eqn:bayesprop}
   \pi_{\Statex_k \given \Meas_{1:k}}(\state \given \meas_{1:k}) \propto \pi_{\Meas_{k} \given \Statex_k}(\meas_{k} \given \state)  \pi_{\Statex_k \given \Meas_{1:k-1}}(\state \given \meas_{1:k-1}).
\end{equation}
This equation provides us with a means of assimilating a new observation, $\meas_k$, into our forecast probability distribution for the state. The forecast itself comes from our dynamical model. Under the Markov assumption for the state dynamics, the distribution of the state at time step $k$ depends only on the state at the previous time step $k-1$, i.e., $\pi_{\Statex_k \given \Statex_{0:k-1}} = \pi_{\Statex_k \given \Statex_{k-1}}$. The joint distribution $\pi_{\Statex_{k-1:k}}$ can thus be factorized into this state transition and the posterior distribution at the end of the previous step:
\begin{equation}
    \label{eqn:markov}
    \pi_{\Statex_{k-1:k} \given \Meas_{1:k-1}} = \pi_{\Statex_k \given \Statex_{k-1}} \pi_{\Statex_{k-1}\given \Meas_{1:k-1}}.
\end{equation}
By recursively applying this equation with (\ref{eqn:bayesprop}) substituted for the posterior distribution, it is easy to show that
\begin{equation}
\label{eqn:bayesprop2}
   \pi_{\Statex_{0:k} \given \Meas_{1:k}} \propto
   \pi_{\Statex_0} \prod_{i=1}^{k}\pi_{\Meas_{i} \given \Statex_i}  \pi_{\Statex_{i} \given \Statex_{i-1}}.
\end{equation}
This equation is central in Bayesian inference since it justifies the use of sequential methods to estimate the posterior distribution. We can update our previous estimate with new observations sequentially without having to restart the calculation at every time step.

The prior distribution $\pi_{\Statex_{k}|\Meas_{1:k-1}}$ can be obtained by marginalizing (\ref{eqn:markov}) over all the possible realizations of the state at time $k-1$: \begin{equation}
\label{eqn:chapman}
    \pi_{\Statex_{k}\given \Meas_{1:k-1}} = \int_{\Statex_k} \pi_{\Statex_{k}\given \Statex_{k-1}}\pi_{\Statex_{k-1} \given \Meas_{1:k-1}} d\Statex_{k-1}.
\end{equation}
This equation is called the Chapman--Kolmogorov equation and corresponds to a direct integration of the state transition kernel.


Thus, Bayes' theorem (\ref{eqn:bayesprop}) coupled with equation (\ref{eqn:chapman}) provides us with an elegant update rule for incorporating new measurements into the probability distribution, and thus, improving our estimate of the mean and uncertainty of the state. It requires a time-discretized dynamical model (\ref{eqn:dynamic}) for the state transition, $\pi_{\Statex_k|\Statex_{k-1}}$, and an observational model (\ref{eqn:obs}) for the likelihood $\pi_{\Meas_k|\Statex_{k}}$.

However, there are two primary challenges with applying this in the contexts of interest here. First, we do not generally know the forms of distributions involved in any of these formulas. In what follows in this paper, we will make the typical assumption that all errors are drawn from Gaussian distributions with zero mean, but there are good reasons to doubt that this is reasonable in the present nonlinear context of a vortex model. Second, for high-dimensional problems, the integration of the Chapman--Kolmogorov equation (\ref{eqn:chapman}) is intractable. Indeed, here, as in many physics contexts, the forecast step corresponds to the time advancement of a partial differential equation (the Euler equations), so it is infeasible that we could advance such an equation over all possible values of the state. In our case, it is a time step of the discrete vortex model---lower dimensional than the discretized Navier--Stokes system, to be sure, but still containing tens to hundreds of degrees of freedom. 

Both of these challenges motivate our use of ensemble filtering methods in this work to sequentially estimate the posterior distribution. We use a set of $M$ {\em particles} $\ensemble{\state^1, \state^2, \cdots, \state^M}$ sampled from the state distribution $\pi_{\Statex_{k-1}\given \Meas_{1:k-1}}$ and we aim to construct a particle approximation of the posterior distribution $\pi_{\Statex_{k} \given \Meas_{1:k}}$. It should be stressed that the term ``particle'' is used here in its usual sense in stochastic estimation as a member of the ensemble; it does not denote a vortex particle, a term that we avoid in this paper in favor of ``vortex element'' or ``vortex blob''.

To construct this particle approximation of the posterior distribution, we perform the following two steps: a forecast step and an analysis step. In the forecast step, each particle $\state^i$ is propagated through the dynamical equation (\ref{eqn:dyn_basic}) (the vortex model) and randomly perturbed with noise (called additive covariance inflation, discussed below) to form samples from the prior distribution $\pi_{\Statex_{k} \given \Meas_{1:k-1} = \meas_{1:k-1}}$. This ensemble forecast step constitutes a Monte Carlo approximation of the Chapman--Kolmogorov equation (\ref{eqn:chapman}). Given this sampling $\ensemble{\state^i}$ from the prior distribution, one can easily create a sampling $\ensemble{\meas^i}$ from the likelihood distribution $\pi_{\Meas_{k} \given \Statex_{k}}$ by evaluating the observation equation (\ref{eqn:obs_basic}) (the Bernoulli equation) for each ensemble member $\state^i, i = 1,\ldots,M$.

The analysis step then updates the set of ensemble members by assimilating the new realization $\meas^\star_{k}$ of the observation variable $\Meas_{k}$; in this paper, $\meas^\star_{k}$ represents a new pressure difference measured from the truth system, a high-fidelity Navier--Stokes simulation. With the probability distributions unknown, this would require that we estimate the posterior distribution via Bayes' theorem (\ref{eqn:bayesprop}) given the finite set of samples $\ensemble{\state^i, \meas^i}$ from the joint distribution formed by the prior and likelihood. However, the task is made considerably simpler if we assume that the distributions are Gaussian. Under that assumption, and with knowledge of the prior mean and covariance matrix, then Bayes' theorem leads naturally to the analysis step of the classical Kalman filter \cite{kalman1960new}. In the particle approximation, in which we estimate this mean and covariance from the finite ensemble statistics that emerge from the forecast, we obtain the ensemble Kalman filter \cite{asch2016data}. It should be noted, however, that since the underlying distributions are likely non-Gaussian, {\em our EnKF framework is not expected to converge in the limit of large ensemble size}. In the following sections we will present two forms of this method, but first we present some important notation.



\subsection{Additional notation}

We will use the superscript $f$ to denote prior, forecast quantities and the superscript $a$ to denote analysis, posterior quantities. In particular, the exact prior mean is $\state_{k}^f = \E{\Statex_{k} \given \Meas_{1:k-1} = \meas_{1:k-1}} = \E{\Statex_{k} \given  \meas_{1:k-1}}$ and exact prior 
covariance is $\Pf_{k} = \cov{\Statex_{k} \given \meas_{1:k-1}}$. The posterior mean and covariance are given by $\state_{k}^a = \E{\Statex_{k} \given  \meas_{1:k}}$ and $\Pa_{k} = \cov{\Statex_{k} \given  \meas_{1:k}}$, respectively.

Given an ensemble $(\state^i)$ of size $M$, we define the ensemble matrix $\X \in \real^{n\times M}$ as:
\begin{equation}
\label{eqn:ensemble}
    \X = \left[\state^1, \state^2, \ldots, \state^M  \right].
\end{equation}
We use an overbar to denote statistics obtained from the ensemble, such as the sample mean and covariance:
\begin{equation}
\label{eqn:sample}
\statebar=\frac{1}{M} \sum_{i=1}^{M} \state^{i},
\quad \Pbar=\frac{1}{M-1} \sum_{i=1}^{M} \left(\state^{i} - \statebar \right) \left(\state^{i}-\statebar \right)^{\top},
\end{equation}
where $\top$ denotes the transpose operator. We define the anomaly matrix $\Xp \in \real^{n \times M}$ of an ensemble as
\begin{equation}
\label{eqn:anomaly}
\Xp = \frac{1}{\sqrt{M-1}}[\state^1 -\statebar, \state^2 -\statebar, \ldots, \state^M -\statebar].
\end{equation}
The anomaly matrix obviously has zero mean. Using $\one$ to denote the vector of ones of length $M$, we can conveniently define the sample mean, anomaly matrix, and sample covariance from the ensemble matrix:
\begin{equation}
\label{eqn:relationensemble}
\statebar = \frac{1}{M}\X \one , \quad \Xp = \frac{1}{\sqrt{M-1}}\X(\id - \one \one^{\top}/M), \quad \Pbar = \Xp \Xp^{\top}.
\end{equation}
Finally, we have the relation between the ensemble matrix and the anomaly matrix:
\begin{equation}
\label{eqn:split}
    \X = \Xbar + \sqrt{M-1} \Xp
\end{equation}
where  $\Xbar = \statebar \one^{\top} = \left[\statebar, \statebar, \ldots, \statebar \right] \in \real^{n\times M}$. It is easy to verify that $\one^\top \one = M$, $\Xp \one = 0$, and $\Xbar\one/M = \statebar$.

\subsection{\label{sub:stochastic}The stochastic ensemble Kalman filter}
The application of the stochastic ensemble Kalman filter for aerodynamic flow estimation was already presented already in \cite{da2018ensemble,darakananda2018data}, but we review it here in order to identify an important deficiency that motivates our current treatment. The forecast step of the \senkf{}---and, indeed, of all forms of the EnKF---is given by applying (\ref{eqn:dynamic}) to each ensemble member:
\begin{equation}
    \label{eqn:forecast_step}
    \state^{f,i}_k = \dyn(\state^{a,i}_{k-1}) + \sampledyn_{k}^i \quad \mbox{for } i=1, \ldots, M,
\end{equation}
where the random process noise $\sampledyn_{k}^i$ is drawn from the distribution $\sampledyn_{k}^i \sim \pi_{\BB{\mathsf{Q}_k}}$. It should be noted that this process noise is not inherently part of the dynamical model and must be explicitly introduced; this is discussed in Section~\ref{sub:regenkf}.

In order to simplify notation, we drop the time dependence subscript of the variables in the rest of this section, since the analysis step does not involve time propagation. The analysis step of the \senkf{} seeks a linear transformation of the form:
\begin{equation}
    \label{eqn:stochupdate}
    \stateai =  \statefi + \K (\meas^\star + \sampleobs^i - \obs(\statefi)), \quad \mbox{for } i=1,\ldots, M,
\end{equation}
where $\statefi$ and $\stateai$ are the $i$th prior and posterior ensemble member, respectively; $\sampleobs^i$ is drawn from the measurement noise distribution $\pi_{\noiseobs}$; and $\meas^\star$ is the realization of the observation variable $\Meas$ at the current assimilation time. The expression $\meas^\star + \sampleobs^i - \obs(\statefi)$ is called the innovation for the $i$th ensemble member.

The algorithm of the \senkf{} is provided in Algorithm~\ref{algo:senkf}. The original analysis step derived by Evensen \cite{evensen1994sequential} did not include the measurement noise term, but this was corrected by Burgers et al. \cite{burgers1998analysis} to reflect that $\meas^\star$ is drawn from a random distribution and to avoid some spurious correlations within the ensemble. The matrix $\K \in \real^{n \times d}$ is called the \textit{Kalman gain} and is identical to its form in the standard Kalman filter, derived to minimize the trace of the posterior covariance matrix $\Pa$ (as we will discuss further below): 
\begin{equation}
\label{eqn:kalman gain}
\K = \Pf \Obs^\top (\Obs \Pf \Obs^\top + \covobs)^{-1}
\end{equation}
with $\covobs$ the covariance matrix of the measurement noise and $\Obs$ the tangent linear of the observation operator. The exact prior covariance $\Pf$ is approximated by the sample prior covariance $\Pbarf$ formed from the prior ensemble $\ensemble{\statefi}$ with (\ref{eqn:relationensemble}), $\Pbarf = \Xpf \Xpf^{\top}$; similarly, the posterior covariance is approximated by $\Pbara = \Xpa \Xpa^{\top}$. 

The Kalman gain suggests that we must calculate the tangent linear of the observation operator. Evensen  \cite{evensen1994sequential} proposed a technique called implicit linearization that approximates $\Pf \Obs$ and $\Obs \Pf \Obs^\top$ given the prior ensemble. First, we construct $\measbarf$ the mean of the observations $\ensemble{\obs(\statefi)}$ for $i=1,\ldots, M$:
\begin{equation}
\label{eq:measbarf}
\measbarf =\frac{1}{M} \sum_{i=1}^{M} \obs \left(\statefi \right).
\end{equation}
Let us then define the innovation anomaly matrix $\Ypf$, with $i$th column given by
\begin{equation}
\label{eqn:obsanomaly}
\Ypfi = \frac{\obs(\statefi) - \measbarf - \sampleobs^i + \overline{\sampleobs}}{\sqrt{M-1}}, \mbox{ for } i=1, \ldots, M,
\end{equation}
and with $\overline{\sampleobs}$ the sample mean of $\ensemble{\sampleobs^i}$. It should be noted that this mean is itself a random number whose expected value is 0.

The key ingredient of the technique is to make the following approximation:
\begin{equation}
\Obs(\statefi- \statebarf) \simeq \obs(\statefi)-\measbarf.   
\end{equation}
Using this approximation, it is easy to show that $\Pf \Obs^\top$ and $\Obs \Pf \Obs^\top$ are approximated by $\Xpf \Ypf^\top$ and $\Ypf \Ypf^\top - \covobs$, respectively \cite{asch2016data}, and thus, the Kalman gain (\ref{eqn:kalman gain}) takes the simple form
\begin{equation}
\label{eqn:kalmanfact}
    \K = \Xpf \Ypf^\top \left( \Ypf \Ypf^\top\right)^{-1}.
\end{equation}


To gain more insights on this formulation of the filter, we can rewrite the analysis update (\ref{eqn:stochupdate}) with ensemble matrix notations. From (\ref{eqn:split}), the update of each posterior ensemble member is equivalent to a separate update of the posterior mean $\statebara$ and the posterior anomaly matrix $\Xpa$. From the linear analysis step (\ref{eqn:stochupdate}), the posterior anomaly matrix $\Xpa$ is updated according to (\cite{asch2016data}):
\begin{equation}
\label{eqn:stochupdatemat}
\Xpa = \Xpf -\K \Ypf;
\end{equation}
the update equation for the posterior mean state $\statebara$ is obtained by taking the expectation of the analysis step (\ref{eqn:stochupdate}):
\begin{equation}
    \label{eqn:etkfmean}
    \statebara = \statebarf + \K(\meas^\star - \measbarf),
\end{equation}
where $\measbarf$ is defined in (\ref{eq:measbarf}). Equation (\ref{eqn:etkfmean}) is the exact Kalman update equation.

Our derivations thus far have omitted the fact that our actual ensemble is finite sized. Because of the assumed linear form of the analysis step, equation (\ref{eqn:etkfmean}) is reproduced almost exactly for finite-sized ensembles, except for the addition of a lingering sample mean of the measurement noise, $\overline{\sampleobs}$, in parentheses. However, a more problematic discrepancy lies in the relationship between the posterior and prior covariance matrices. For obtaining this relationship, let us first denote by $\covobsp$ the anomaly matrix of the measurement noise $\covobspi = (\sampleobs^i - \overline{\sampleobs})/\sqrt{M-1}$; the product $\covobsp \covobsp^\top$ approximates the measurement covariance $\covobs$. 
Then, from the update equation (\ref{eqn:stochupdatemat}) and the definitions of the covariances and innovation anomaly matrix, we can construct the equation for the posterior covariance matrix: 
\begin{equation}
\label{eqn:covstoch}
    \Pbara = (\id_n  - \K\Obs)\Pbarf(\id_n  - \K\Obs)^{\top} + \K \covobsp \covobsp^{\top} \K^{\top} + (\id_n -\K \Obs)\Xpf \covobsp^{\top} \K^{\top} + \K \covobsp \Xpf^{\top}(\id_n - \K \Obs)^\top.
\end{equation}

By taking the expectation of this matrix over the measurement noise process and then minimizing its trace over $\K$, we recover the classical Kalman gain (\ref{eqn:kalman gain}) and a simple form of the expected value of the posterior covariance:
\begin{equation}
    \E{\Pbara} = (\id_n - \K \Obs) \Pbarf.
\end{equation}
Indeed, the classical Kalman filter adopts this expected value in order to propagate the covariance matrix, $\Pa = (\id_n - \K \Obs) \Pf$. However, the finite sample covariance $\Pbara = \Xpa \Xpa^{\top}$ does not reproduce this ideal relation except in the limit of infinite ensemble sizes. Rather, the covariance of the finite ensemble only strictly satisfies (\ref{eqn:covstoch}), due to spurious correlations between the forecast anomalies and observation noise, $\Xpf\covobsp^\top$. Thus, the stochastic analysis step of the \senkf{} introduces error that degrades the performance of the ensemble Kalman filter. An alternative form of the EnKF will be presented in the next section that addresses this issue.

\subsection{\label{sub:etkf}The ensemble transform Kalman filter}


In the previous section, it was shown that finite ensembles do not reproduce the expected value of the posterior covariance matrix $\Pbara$, and that this can cause the performance of the \senkf{} to degrade. To circumvent the issue, Bishop et al.~\cite{bishop2001adaptive}  developed the \textit{ensemble transform Kalman filter (\etkf{})} that exactly reproduces the ideal propagation equation for the covariance. This has been shown to give better performance in other applications \citep{raanes2015improvements, asch2016data}. The \etkf{} belongs to a more general class of ensemble Kalman filters---called \textit{deterministic ensemble Kalman filters}---that verify exactly the correct propagation equation through various analysis schemes.

To generate a posterior anomaly matrix $\Xpa$, one can start from the desired propagation equation for the covariance and use the factorization of the sample covariance matrix (\ref{eqn:relationensemble}):
\begin{equation}
    \label{eqn:etkfcov}
    \begin{aligned}
    \Pbara  & = \Xpa \Xpa^\top = (\id_n - \K \Obs) \Pbarf
\end{aligned}
\end{equation}
The right-hand side of this equation can also be factorized when $\Pbarf$ is replaced with its own factorization and the ensemble expression (\ref{eqn:kalmanfact}) of the Kalman gain is introduced. In the following, it should be noted the innovation anomaly matrix $\Ypf$ is defined as in (\ref{eqn:obsanomaly}), but now without the observation noise, so that $\Ypf = \Obs \Xpf$:
\begin{equation} 
\label{eqn:etkfcov2}
\begin{aligned}
    \Xpa \Xpa^\top & = (\id_n - \K \Obs) \Xpf \Xpf^\top\\
    & = (\id_n - \Xpf \Ypf^\top(\Ypf \Ypf^{\top} + \covobs)^{-1}\Obs) \Xpf  \Xpf^\top\\
    & = \Xpf  (\id_M - \Ypf^\top(\Ypf \Ypf^{\top} + \covobs)^{-1}\Ypf) \Xpf^\top\\
    & = \Xpf  \G \Xpf^\top.
    \end{aligned}
\end{equation}
Matrix $\G \in \real^{M \times M}$ is symmetric positive definite. In order to produce an analysis equation for the posterior anomaly matrix, we seek a square-root factorization of $\G$. In other words, we look for a matrix $\sqG$ such that $\G = \sqG \sqG^\top = \sqG^\top \sqG$. For a positive-definite matrix, the square-root decomposition exists but is not unique. Indeed for an arbitrary orthogonal matrix $\U \in \BB{O}(M)$---i.e., any matrix such that $\U \U^\top = \U^\top \U = \id_M$---then $\sqG\U$ is also a square-root of $\G$.

Therefore, the analysis update of the anomaly can be written as a right linear transformation:
\begin{equation}
\label{eqn:etkf_anomaly}
  \Xpa =   \Xpf  \sqG \U
\end{equation}
with some choice of $\U \in \BB{O}(M)$, discussed below. Furthermore, the analysis update of the sample mean (\ref{eqn:stochupdatemat}) can be written with the help of (\ref{eqn:kalmanfact}) as
\begin{equation}
\label{eqn:statea}
    \statebara = \statebarf + \Xpf \Ypf^\top (\Ypf \Ypf^{\top} + \covobs)^{-1} \BB{\delta},
\end{equation}
where $\BB{\delta} = \meas^\star - \obs(\statebarf)$ is the mean innovation. Equations (\ref{eqn:etkf_anomaly}) and (\ref{eqn:statea}) are used together, through (\ref{eqn:split}), to update the ensemble. 

It is notable that this analysis step assembles both the posterior anomaly and the update to the mean from the columns of $\Xpf$. In other words, the analysis is said to be performed in the \textit{ensemble space}: the update to the state vector is a linear combination of the deviations among the forecast ensemble members from the mean. In our case, for example, the position of a certain vortex element will be updated with a linear combination of the ensemble deviations of that element's position from the ensemble mean, after the ensemble has been advanced by the vortex model. This emphasizes the importance of such variance among ensemble members: without it, any discrepancy $\BB{\delta}$ between the new observation and the predicted observation is simply ignored.

The form of this update is an attractive property of the \etkf{} since the ensemble size $M$---which is on the order of 100---is typically similar to the size of the state $n$ encountered in the vortex model, and certainly much smaller than the state vector associated with a CFD simulation. Using Sherman\textendash Morrison\textendash Woodbury identities, the formula for $\sqG$ can be simplified:
\begin{equation}
    \sqG = {(\id_M - \Ypf^\top(\Ypf \Ypf^{\top} + \covobs)^{-1}\Ypf)}^{1/2} = {(\id_M +  \Ypf^{\top}\covobs^{-1} \Ypf)}^{-1/2}.
\end{equation}

In this work, we choose the symmetric square-root factorization, $\G = \sqG \sqG$, among the different possibilities. Given an eigendecomposition $\G = \BB{R} \BB{\Sigma} \BB{R}^\top$, the symmetric square root $\sqG$ is given by $\BB{R} \BB{\Sigma}^{1/2} \BB{R}^\top$, with $\BB{\Sigma}^{1/2}$ the entry-wise positive square root of the diagonal matrix $\BB{\Sigma}$. Several studies \citep{livings2008unbiased,evensen2009ensemble, raanes2015improvements} have shown that the symmetric root has useful properties, particularly that it does not introduce a spurious mean in the posterior anomaly matrix: $\G \one = \one$, so $\one$ is an eigenvector of $\G$ with unit eigenvalue; thus, this is also true of $\sqG$, so $\Xpf \sqG \one = \Xpf \one = \zero$. The overall construction of the size $M$ square-root $\sqG$ is computationally inexpensive.

Sakov and Oke~\cite{sakov2008deterministic} have found that the choice of the symmetric square root leads consistently to better performance. However, for large ensemble, the symmetric square root can lead to the creation of outliers in the ensemble. One can prevent the appearance of these spurious members by right multiplying $\sqG$ by a random orthogonal matrix $\U$ from time to time. To avoid bias in the posterior anomaly matrix, we must also ensure that $\Xpa$ has zero mean. To do so, $\U$ must preserve the mean, equivalent to requiring that $\U \one = \one$. Some authors \citep{nerger2012unification, todter2015second} have have proposed a scheme based on sampling of the standard normal distribution and use of Householder reflections to construct such mean-preserving random rotations; this algorithm is presented in Algorithm~\ref{algo:rand}. For the other assimilation steps, $\U$ is simply the identity $\id_M$. The overall \etkf{} is summarized in Algorithm~\ref{algo:etkf}.

\subsection{\label{sub:regenkf}Covariance inflation} 

In the different flavors of the EnKF, one seeks to estimate the posterior covariance matrix of a state variable of large dimension $n$ with only a few ensemble members, $M \sim 100$. Therefore, the posterior covariance is usually underestimated with large sampling errors. This rank deficiency typically leads to long-range spurious correlations within the covariance matrix. Over multiple assimilation cycles, the Kalman gain will decrease to zero and the measurements will no longer be assimilated into the ensemble; this is called \textit{filter divergence}. However, efficient regularization techniques such as the \textit{covariance inflation} and \textit{localization} have been developed to mitigate these sampling errors. These techniques underpin the success of the ensemble Kalman filters in high-dimensional filtering problems. Localization is not implemented in this work but will be discussed in  section \ref{sec:results} to circumvent certain flaws of our flow estimator and will be investigated in future work. 

Covariance inflation is typically applied between the forescast and analysis steps. The \textit{multiplicative inflation} increases the spread of the ensemble about the sample mean,  by rescaling the deviation $\statefi - \statebarf$ by the multiplicative factor $\beta>1$ for each ensemble member:
\begin{equation}
\label{eqn:multiplicative}
    \statefi \xleftarrow{} \statebarf + \beta (\statefi - \statebarf), \mbox{ for } i=1, \ldots, M.
\end{equation}
This form of inflation is equivalent to multiplying the sample prior covariance $\Pbarf$ by $\beta^2$; the best results are typically found when $\beta$ is set to $1.01$ and $1.03$ for the \senkf{} and \etkf{}, respectively.

\textit{Additive inflation}, in contrast, accounts for the process noise that is inherently missing from the dynamical model; such noise is critical for ensuring that the ensemble maintains a variance among its members. This technique adds to each ensemble member a sample from a Gaussian distribution with zero mean and covariance $\BB{Q}$. The additive inflation corresponds to a Tikhonov regularization of the prior covariance matrix:
\begin{equation}
    \Pbarf \xleftarrow{} \Pbarf + \BB{Q}.
\end{equation}

Whitaker and Hamill~\cite{whitaker2012evaluating} proposed to combine the additive and multiplicative inflations. Indeed, Anderson and Anderson \cite{anderson1999monte} have shown that additive and multiplicative inflation account for distinct kind of errors. The multiplicative inflation tends to account for sampling errors due to the small ensemble size, while the additive inflation tends to correct intrinsic errors of the model. Multiplicative inflation only requires the tuning of one parameter. Adjusting additive inflation requires tuning as many parameters as the dimension of the state. This requires more work and a deeper understanding of the dynamical system. Raanes \cite{raanes2015improvements} showed that the additive inflation is a robust technique to account for model errors, more complex inflation schemes can be constructed to reduce further model errors which are not considered in this study \citep{whitaker2012evaluating, raanes2015improvements}.


\subsection{Flow estimation with an ensemble of aggregated vortex models}
\label{sub:details}

With the aggregated vortex model summarized in Section~\ref{sec:vortex} and the details of the ensemble filtering methods discussed above, we are now in a position to summarize the overall method. A schematic of this method is presented in Fig.~\ref{fig:algo_vortex}. In the forecast, an ensemble of aggregated vortex models are advanced by one time step. At the end of this forecast, multiplicative and additive covariance inflation are applied to each member of the ensemble. The analysis step is then carried out, either by applying Algorithm~\ref{algo:senkf} for the \senkf{} or Algorithm~\ref{algo:etkf} for the \etkf{}.

It is useful to interpret the method physically. Before we discuss the physical roles of each step of the filter, it is important to understand the effect of the ensemble and its propagation on the vorticity field. To demonstrate this effect in the clearest manner, let us assume that the $q$ vortex elements in each ensemble member are singular: point vortices rather than blobs. Their blob form is only used to regularize the Biot--Savart interactions between them, and our discussion here focuses only on the interpretation of the vorticity field itself.

We can then write the vorticity field at location $\BB{y}$ and time step $k$ as a function of the random state vector $\Statex_{k}$, whose components (aside from $\lespc{}$) constitute the strengths and positions of the singular elements:
\begin{equation}
    \omega(\BB{y},\Statex_{k}) = \sum_{j=1}^{q} \Gamma^{j}_{k} \delta(\BB{y} - \BB{y}_{k}^{j}),
\end{equation}
where $\delta$ is the Dirac delta function. The expected value of the vorticity field at the end of time step $k$ is given (ideally, for infinite ensemble) in terms of the posterior distribution function
\begin{equation}
    \E{\omega(\BB{y},\Statex_k)} = \int \omega(\BB{y},\state_k) \pi_{\Statex_k} (\state_k)\,\mathrm{d}\state_k.
\end{equation}
(For simplicity of notation, we have omitted the fact that the distribution $\pi_{\Statex_k}$ is conditioned on the observations made thus far.) Under our Gaussian assumption, it is particularly easy to evaluate these integrals, and we arrive at
\begin{equation}
    \E{\omega(\BB{y},\Statex_k)} = \sum_{j=1}^q \frac{\overline{\Gamma}_k^j}{2\pi |\BB{P}^{y_j}_k |^{1/2}} \exp \left(-\frac{1}{2}(\BB{y}-\overline{\BB{y}}^{j}_k)^\top {\BB{P}^{y_j}_k}^{-1} (\BB{y}-\overline{\BB{y}}^{j}_k)\right),
\end{equation}
where $|\cdot|$ denotes determinant; $\overline{\Gamma}_k^j$ and $\overline{\BB{y}}^{j}_k$ denote, respectively, the mean circulation and position of vortex element $j$ at time step $k$; and $\BB{P}^{y_j}_k$ is the $2\times 2$ covariance submatrix associated with the position of vortex element $j$ at step $k$.

In other words, we can interpret each vortex element's uncertainty as defining an elliptically-shaped region, centered at its mean location and endowed with its mean strength. The behavior of this elliptical region's shape over time is determined by the combined influences of three processes in the filter: the forecast, the inflation, and the analysis. 

The role of the forecast step is straightforward: it constitutes an inviscid (i.e., advective) advancement of the vortex elements by one step and the creation of new vorticity to satisfy modeled edge conditions. The ensemble of such inviscid models establishes a set of slightly different displacements of each vortex element. Since we interpret this ensemble as approximating a Gaussian distribution both before and after the forecast, the set of displacements of each vortex define a constrained transformation of the vortex's ellipse: a net advection of the center and a stretching and rotation of its shape. 

The inflation step imposes two influences on each region's shape. The multiplicative inflation stretches the ellipse uniformly by a small fraction in every direction. The additive inflation, on the other hand, comprises a single step of a random walk. It is well known that a random walk, applied over a large number of steps, approaches a Wiener process, and the associated probability distribution satisfies a linear diffusion equation \cite{pavliotis2014stochastic}. This diffusion causes the elliptical region to spread, reminiscent of core spreading in vortex methods \cite{leonard:3j}. To simulate diffusion of viscosity $\nu$, the random step is chosen from a normal distribution with standard deviation $\sqrt{2\nu \Delta t}$.

In the context here, the additive inflation occurs among other processes, so its interpretation is less clear. Chorin's random vortex method \cite{chorin1973numerical}, utilizing a large number of overlapping vortex blobs undergoing random walks, is known to converge to the solution of the Navier--Stokes equation as $q^{-1/2} \log q$, where $q$ is the number of blobs  \cite{long1988convergence}. In the EnKF context, it is possible to identify a stochastic differential equation that asymptotically describes the forecast and inflation steps in the limit of large ensemble \cite{Evensen2003TheImplementation}. We leave a rigorous interpretation of the ensemble of vortex models in this manner for future work. We merely observe that the variance of the distribution from which we select the additive inflation parameter loosely specifies the viscosity of a diffusion process. However, we do not attempt to match this effective viscosity to the actual viscosity in the truth system; rather, we tune the additive inflation to balance the trust between the forecast and analysis steps.

Finally, in order to reconcile the new observations made in this step---namely, the pressures measured on the surface of the wing---with the pressures predicted via the Bernoulli equation, the vortex elements' positions and strengths (and \lespc{}) need to be adjusted. The EnKF analysis step provides this adjustment by assembling a minimum least-squares solution to the problem $\obs(\statebara) = \meas^\star$, subject to measurement noise $\covobs$; the minimization is regularized by the forecast $\statebarf$, with its associated covariance $\Pbarf$, ensuring that the new state vector does not stray far from its forecast. Mathematically, this problem is expressed as \cite{da2020flow, Law2015}: 
\begin{equation}
    \statebara = \argmin_{\state \in \BB{\mathsf{R}}^n} \frac{1}{2}(\meas^\star - \obs(\state))^\top \covobs^{-1}(\meas^\star - \obs(\state)) + \frac{1}{2\beta}(\state - \statebarf)^\top {\Pbarf}^{-1}(\state - \statebarf),
\end{equation}
with $\beta$ the multiplicative covariance inflation factor. The analysis step causes the covariance to shrink, according to (\ref{eqn:covstoch}). The elliptical region associated with each vortex element is translated, stretched, and rotated by the analysis step, but its area shrinks.

It should be noted that the dimension of the state vector, $n$, changes with each time step of this filtering process: it increases typically by six (three per newly released blob originating from each edge), and occasionally decreases as elements of zero strength are eliminated. There are no inherent restrictions in the EnKF on changes to the dimension of the state. However, in order to keep a consistent state dimension, i.e., the same number of blobs across the different ensemble members, the vortex elements whose circulations are aggregated into another element are not removed, but rather, are simply assigned zero circulation. Similarly, even if the \lesp{} does not exceed the current estimate of the \lespc{}, a new vortex with zero circulation is still introduced. If the blob has zero circulation across all ensemble members, then it is removed from the state vector.

We discretize the vortex model with the forward Euler time scheme with time step $\Delta t^\star = 0.01$. The blob radius $\delta$ (normalized by $c$) is set to $5\times 10^{-3}$ and $9\times 10^{-3}$ for the \senkf{} and the \etkf{}, respectively. For the aggregated vortex model, the vortex elements are mostly isolated and their interactions require little regularization. The non-zero blob radius primarily regularizes the interactions of the vortex elements soon after their release from the edges. Throughout this study, we use an ensemble of size $M = 50$.  At the initial time, no vortices are present in the state vector of the vortex model. The ensemble is initiated with random samples for the \lespc{} drawn from $\N(0.5, 0.1)$, i.e., a normal distribution with mean $0.5$ and covariance $0.1$.

We use a different set of inflation parameters for the \senkf{} and the \etkf{}, obtained by trial and error:
\begin{itemize}
    \item A multiplicative inflation $\beta = 1.01, 1.028$ for the \senkf{} and \etkf{}, respectively.
    \item An additive inflation for the position of the vortices drawn from $\N(0, 1 \times 10^{-5})$ (normalized by $c$)
    \item An additive inflation for the strength of the vortices drawn from $\N(0, 1 \times 10^{-3}\Delta t^\star)$ (normalized by $Uc$)
    \item An additive inflation for the \lespc{} drawn from $\N(0, 5 \times 10^{-5})$, $\N(0, 8.5 \times 10^{-5})$ for the \senkf{} and \etkf{}, respectively.
    \item The measurement noise $\noiseobs_k$ is drawn from $\N(0, 1 \times 10^{-8})$ (normalized by $\rho U^2$)
\end{itemize}
The \senkf{} and \etkf{} share the same additive inflation parameters for the vortex properties. It is important to note that each of these parameters is chosen in order to balance the trust between the vortex model and the analysis step; lower values, for example, lead to a vortex model that is less responsive to measurement innovation. For example, the relatively larger additive inflation for the \lespc{} makes this parameter more responsive than the vortex parameters. The same parameters are used for all examples in the next section.

\section{\label{sec:results}Results}

In this section, we present the results of flow estimation for two strongly-perturbed flows about an infinitely thin plate at $20^\circ$. In the first case, a sequence of perturbations that mimic pulse actuation is applied near the leading edge of the plate. In the second case, the plate is subject to large scale and coherent perturbations created in the wake of an upstream cylinder. In the discussion of the results, positive and negative vortex elements will refer to vortex elements with positive (counter-clockwise) and negative (clockwise) circulation respectively. 

In this study, the true pressure jump measurements are generated from simulations of a flat plate at Reynolds number 500, carried out with a high-fidelity Navier--Stokes solver, based on the immersed boundary projection method with lattice Green's function \citep{taira2007immersed,  liska2017fast}. It is important to emphasize that the flow perturbations are only present in the truth system, and their effect is only made available to the vortex model through the assimilation of the true pressure measurements. 

Due to the stochastic nature of the filtering problem, results may vary from one simulation to another for the same filter. We estimate the uncertainty in the results in order to draw consistent conclusions about the performance of each filter. For assessment purposes, the results presented in this work have been obtained by running an ensemble of $100$ realizations of the same filter on each flow configuration, with each realization consisting of an application of the EnKF (using an $M = 50$ ensemble of vortex models). From this ensemble of realizations, we construct the sample mean and standard deviation of the quantity of interest for comparison with the truth. If we assume that the results obtained over the different runs follow a Gaussian distribution, the $95\%$ confidence interval can be estimated by considering the plus and minus deviation of twice the standard deviation from the mean. It should be noted that the number of vortex elements---and hence, the dimensionality of the state vector---varies from one realization of the filter to the next, so it is not possible to define a mean state from the ensemble of realizations.

As mentioned in Section~\ref{sec:vortex}, the near encounters between the singular vortex elements and the plate can lead to spurious errors in the estimate of the pressure distribution, and hence, in the prediction of overall normal force on the plate. We use a median filter to remove these spurious spikes from our presented results. At each time step, the current value is replaced by the median value of the $l = 7$ previous time steps. The median filter is causal, so it is directly integrated into the estimation sequence.


\subsection{\label{sub:actuation}Translating plate subjected to pulse actuation disturbances}

\begin{figure}
    \centering
    \includegraphics[width = 0.8\linewidth]{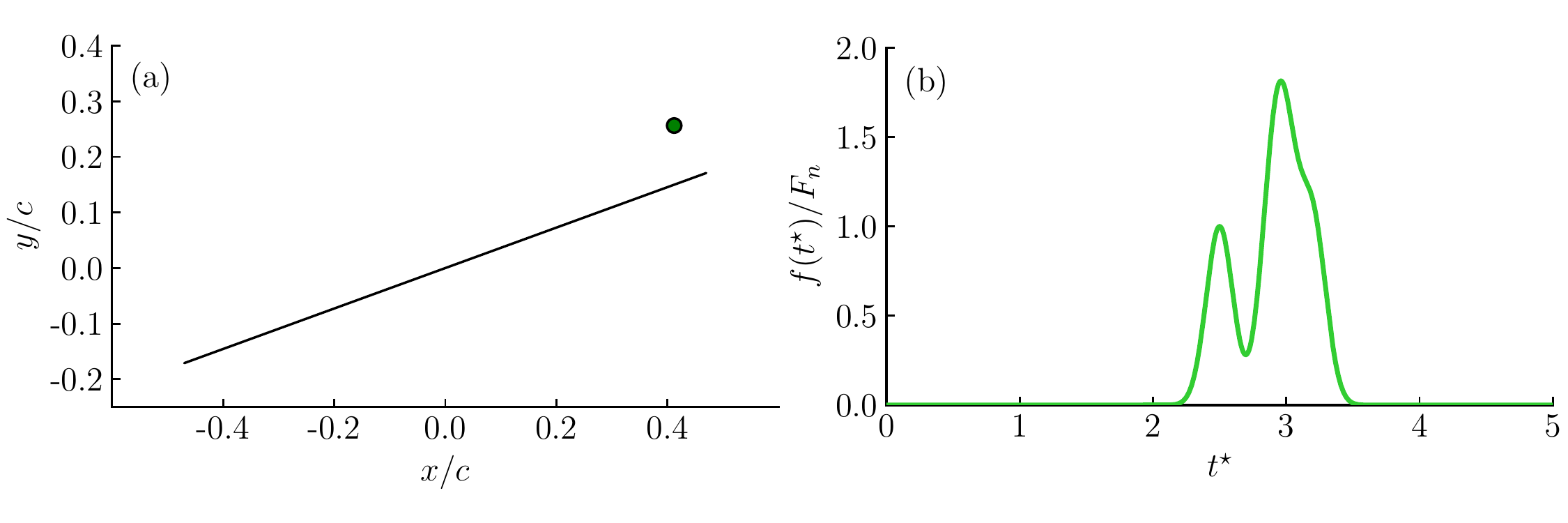}
    \caption{(a): Schematic of the plate at $20^\circ$ at $\ts = 0$ subject to actuation. Green dot depicts the location of the actuation. (b) Time history of the flow actuation.}
    \label{fig:actuation}
\end{figure}

In this section, we assess the data-assimilated vortex model on the response of an impulsively translating plate at $20^\circ$ to disturbances applied near the leading edge. We apply the perturbations at a point $0.475c$ from the centroid along the plate and $0.1c$ above it, as shown in Fig.~\ref{fig:actuation}(a). The pertubations are introduced as a superposition of vertical body force of nominal strength $F_n = 0.03 \rho U^2 c$ at $t^\star = 2.5$, $2.9$, $3.0$ and $3.2$, distributed in Gaussian form in time and space, with temporal standard deviation of $t^\star_{std} = 0.1$ and spatial standard deviation $0.1c$. The vertical body force $f(t^\star)$ is given by:
\begin{equation}
    \frac{f(t^\star)}{F_n} = \N(t^\star;2.5, t^\star_{std}) + \N(t^\star;2.9, t^\star_{std}) + \N(t^\star;3.0, t^\star_{std}) + \N(t^\star;3.2, t^\star_{std}),
\end{equation}
where $\N(t^\star;\mu, \sigma)$ denotes the temporal Gaussian kernel, with mean $\mu$ and standard deviation $\sigma$, evaluated at $t^\star$. Fig.~\ref{fig:actuation}(b) shows the time history of the force actuation. For reference, the gusts considered in Darakananda et al.~\cite{darakananda2018data} were weaker and non-overlapping, applied further upstream of the plate at $t^\star = 3, 4$ with nominal amplitude $F_n = 0.01 \rho U^2 c$. The disturbance considered here is more challenging due to the presence of strong and overlapping perturbations.


\begin{figure}
    \centering
    \includegraphics[width = \linewidth]{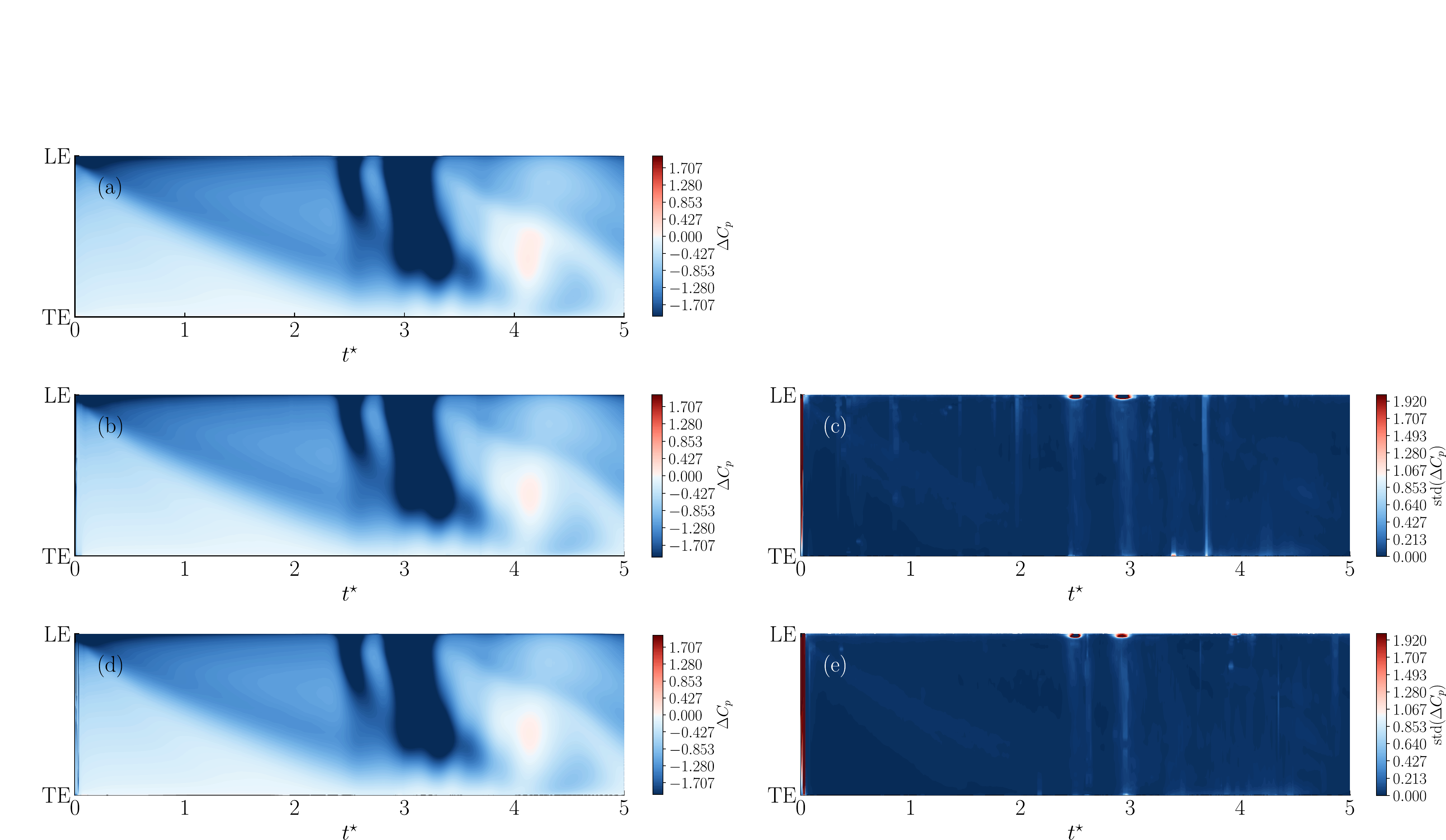}
    \caption{Left column [(a), (b), (d)]: Spatiotemporal map of the pressure coefficient jump for an impulsively translating plate at $20^\circ$ subjected to pulse actuation disturbance from (a) high-fidelity numerical simulation at Reynolds number $500$, and mean over $100$ realizations of an inviscid vortex model with (b) the \senkf{} and (d) the \etkf{}. Right column [(c), (e)]: Spatiotemporal map of standard deviation of the pressure coefficient jump over $100$ realizations for (c) the \senkf{} and (e) the \etkf{}}
    \label{fig:case4_pressure}
\end{figure}

Fig.~\ref{fig:case4_pressure} compares the history of the surface pressure response from the truth with the mean of $100$ realizations of the \senkf{} and the \etkf{}. During the first two convective times, we observe the development of the leading-edge vortex. The successive flow actuation disturbances are easily detected by the distinct regions of strongly negative pressure. The first suction region is due to the disturbance applied at $t^\star = 2.5$, while the second one is due to the superposed response to the flow disturbances centered at $t^\star = 2.9$, $3.0$, and $3.2$. Both filters match well with the true pressure distribution. The uncertainty of each filter is characterized by the sample standard deviation of the pressure, computed over the 100 runs; these are shown in the right column of Fig.~\ref{fig:case4_pressure}. In both filters, two narrow bands of high variance, particularly strong near the leading edge, can be identified at $\ts = 2.5$ and $3.0$. These correspond to the instants of local maxima in the disturbance force. However, the \senkf{} has a higher level of variability in the pressure distribution from one run to another. In particular, there is a significant band of high dispersion for the \senkf{} at around $\ts = 3.8$, and additional smaller bands at other times. The \etkf{} does not exhibit such bands, a direct consequence of its statistically consistent analysis update. 

\begin{figure}
    \centering
    \includegraphics[width = 0.9\textwidth]{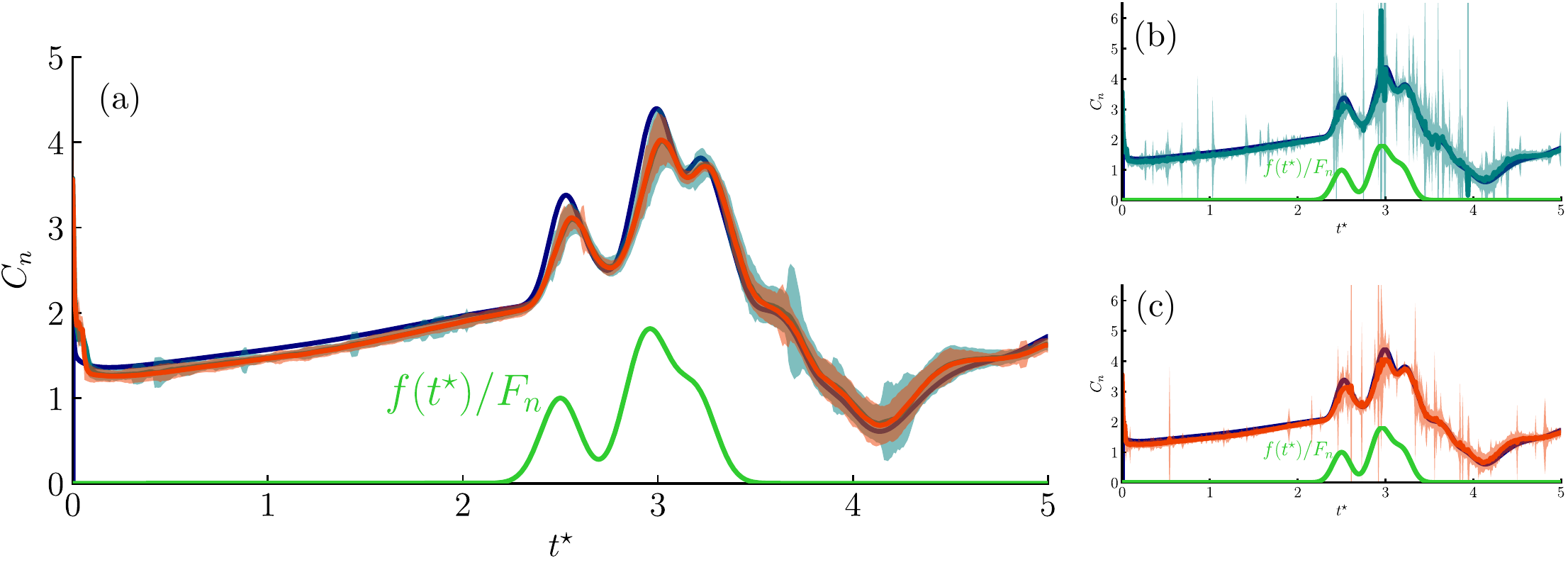}
    \caption{Left panel (a): Normal force coefficient of an impulsively translating plate at $20^\circ$ subject to actuation, from high-fidelity numerical simulation at Reynolds number $500$  \lcolor{ccfd}, mean over $100$ realizations of the inviscid vortex model with \senkf{} \lcolor{cenkf}, mean over $100$ realizations of the inviscid vortex model with  \etkf{} \lcolor{cetkf}. Shaded areas show the $95\%$ confidence interval for the inviscid vortex model with \senkf{} and \etkf{}. Time history of the flow actuation \lcolor{cgust}. Right panel [(b), (c)]: Comparison with the same normal force coefficient from high-fidelity simulation, but without application of the median filter for the \senkf{} (b) and \etkf{} (c).}
    \label{fig:case4_force}
\end{figure}


The normal force coefficient is the integral of this pressure distribution on the plate. The left panel in Fig.~\ref{fig:case4_force} compares the force coefficient obtained from the truth system with the mean of the \senkf{} and the \etkf{} applications over $100$ realizations. The mean force predicted by each filter agrees very well with the true force response. The peaks created by the actuation and the subsequent drop of force around $\ts = 4$ are also well predicted. However, consistent with our observations of the surface pressure data, both filters show variability near the peak disturbances; Fig.~\ref{fig:case4_force} also shows that the peaks are slightly underpredicted.

The \senkf{} exhibits more variability that the \etkf{} at all times, and this higher variability is apparent in the wider uncertainty envelope for the normal force coefficient, particularly after $\ts > 3.5$. To better appreciate the raw behavior of each filter, the right panels in Fig.~\ref{fig:case4_force} show the same results, but without the use of the median filter. The \senkf{}, in Fig.~\ref{fig:case4_force}(b), exhibits frequent spikes throughout the simulation, and particularly so after $\ts > 3.5$. These spikes are not eliminated by the \etkf{}, but they are much weaker and less frequent.

\begin{figure}[thbp]
    \centering
    \includegraphics[width = \linewidth]{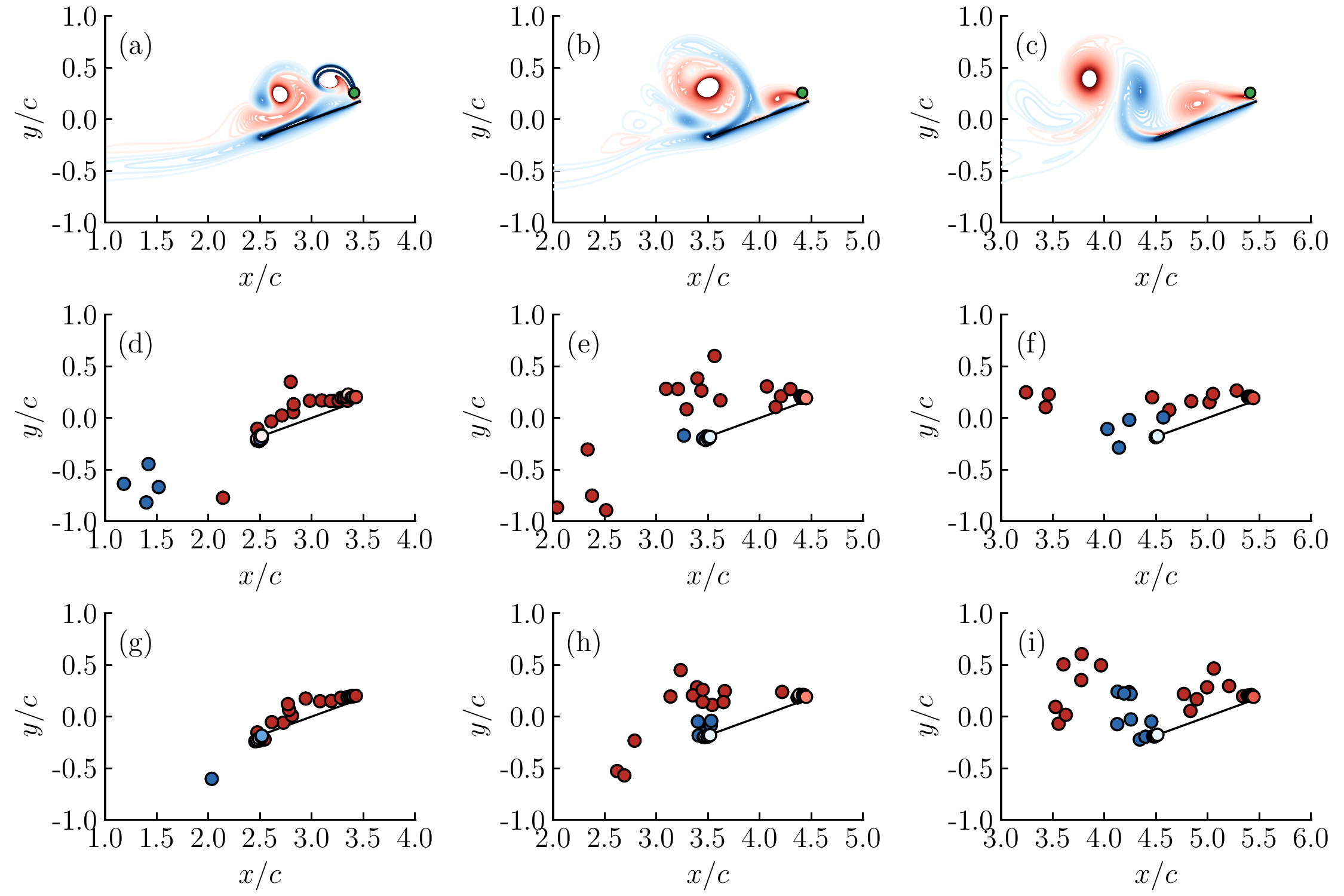}
    \caption{Snapshots of the vorticity distribution at $\ts = 3.0$ (left column), $\ts = 4.0$ (middle column) and $\ts = 5.0$ (right column) for an impulsively translating plate at $20^\circ$ subject to actuation, predicted from [(a)-(c)] high-fidelity numerical simulation at Reynolds number $500$, [(d)-(f)] inviscid vortex model with \senkf{}, and [(g)-(i)] inviscid vortex model with \etkf{}.}
    \label{fig:case4_vorticity}
\end{figure}

\begin{figure}[thbp]
    \centering
    \includegraphics[width = 0.6\textwidth]{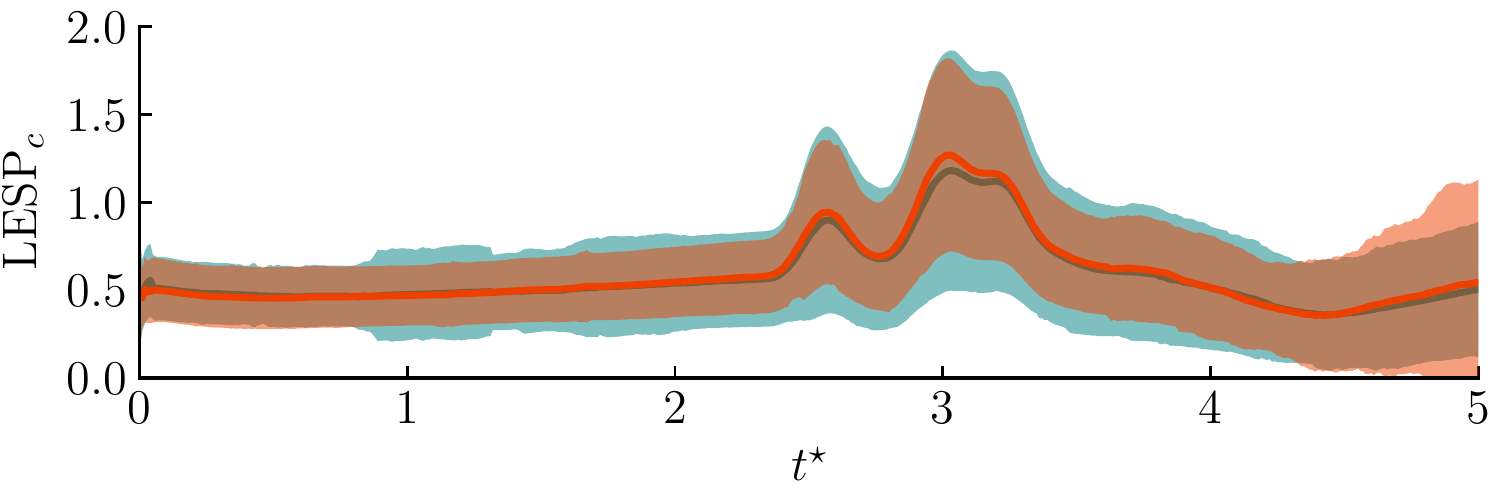}
    \caption{Time history of the ensemble mean value of the \lespc{} of an impulsively translating plate at $20^\circ$ subject to actuation, averaged over $100$ realizations, from the inviscid vortex model with \senkf{} \lcolor{cenkf} and the inviscid vortex model with \etkf{} \lcolor{cetkf}. Shaded areas show the $95\%$ confidence interval for the inviscid vortex model with \senkf{} and \etkf{}.}
    \label{fig:case4_lespc}
\end{figure}

The origins of this high uncertainty are clear when we examine the flow behavior. In Fig.~\ref{fig:case4_vorticity} we show the vorticity distribution from the truth system at three instants,  $\ts = 3.0$, $4.0$, and $5.0$. The models' vorticity is singular, so it is not possible to make a direct comparison; instead, we plot the locations of the elements for one realization of each filter, with the elements' signed circulations represented in colors consistent with the sign of vorticity in the truth data. Most of the large-scale vortex dynamics are captured by the data-assimilated vortex model. Over the first two convective times, the impulsive translation of the plate creates a leading-edge vortex that grows in size and strength. This feature is captured by the continuous release of new vortex elements in both filters. The disturbance applied at $\ts = 2.5$ creates a coherent structure that is rapidly advected along the plate and merges with the initial leading-edge vortex. At around $\ts = 3.0$, the resulting vortex lies slightly above the mid chord of the plate, evident in the left column of Fig.~\ref{fig:case4_vorticity}. Both the \senkf{} and \etkf{} capture this feature with positive vortex elements clustered at a similar position. It is also important to note that the vorticity associated with the disturbance itself, clearly evident at $\ts = 3.0$, is {\em not} observed in the vortex model results. The absence of this disturbance vorticity is by design; only its effect on vortex elements shed from the edge of the plate is captured by the assimilation of pressure data.


Over the time interval $\ts \in [3.0, 4.0]$, the vortex models predict a separation of a cluster of positive vortex elements from the plate with dynamics similar to the truth flow. The middle column of Fig.~\ref{fig:case4_vorticity} compares the results at $\ts = 4.0$. The large coherent structure with positive vorticity is well captured by the inviscid model with both filters, but is more tightly clustered in the case of the \etkf{}. From $\ts = 4.0$ to $5.0$, the vortex is shed into the wake and triggers a large flux of negative vorticity from the trailing edge. The right column of Fig.~\ref{fig:case4_vorticity}, for $\ts = 5.0$, shows a good visual agreement between the truth and the prediction of the \etkf{}. The spatial vortex distribution predicted by the \senkf{} is less representative of the true vorticity distribution, however. Overall, the \etkf{} predicts a more structured and physically consistent spatial distribution of the vortices than the \senkf{}. The coherence of these vortex element clusters is consistent with the narrower uncertainty in the normal force coefficient, and demonstrates a clear advantage of the \etkf{} over the \senkf{} for modeling the flow response to unknown flow perturbations.

In Fig.~\ref{fig:case4_lespc} we compare the time histories of the \lespc{} estimated by the \senkf{} and the \etkf{}. The \lespc{} is constrained to remain positive, and the constraint reverts to the Kutta condition if \lespc{} becomes zero. Before the flow is disturbed, the mean estimate of the \lespc{} stays on a plateau about $0.5$, the mean value in the initial ensemble. This behavior supports the hypothesis of Ramesh et al.~\citep{ramesh2013unsteady, ramesh2014discrete}: the \lespc{} remains constant for a given Reynolds number and airfoil section. The time variation of the imposed disturbance is reflected in a similar variation of the \lespc{}. Indeed, the application of an actuation-like disturbance near the leading edge directly controls the vorticity flux about this edge. Large values of \lespc{} lead to weaker vorticity, temporarily suppressing the flux into the shear layer. The small decay of \lespc{} after the first disturbance increases the vorticity flux, triggering the creation of a new coherent structure that merges with the initial leading-edge vortex about the mid-chord. This leading-edge development is then halted after the next disturbance peak, and the leading-edge vortex is shed. The uncertainty envelopes of the two filters are very similar and tend to grow over time. The width of these envelopes is large, reflecting significant variation in the estimated values of \lespc{} from one realization to the next. This variation indicates a weak physical correlation between this threshold value and the pressure on the plate: this threshold's effect on pressure is only exerted indirectly, through the subsequent release of vorticity. (The \lesp{} itself, in contrast, is more strongly correlated, since shed vortex elements contribute to this value \cite{ramesh2014discrete,eldredgebook}.)


\begin{figure}
    \centering
    \includegraphics[width = 0.6\linewidth]{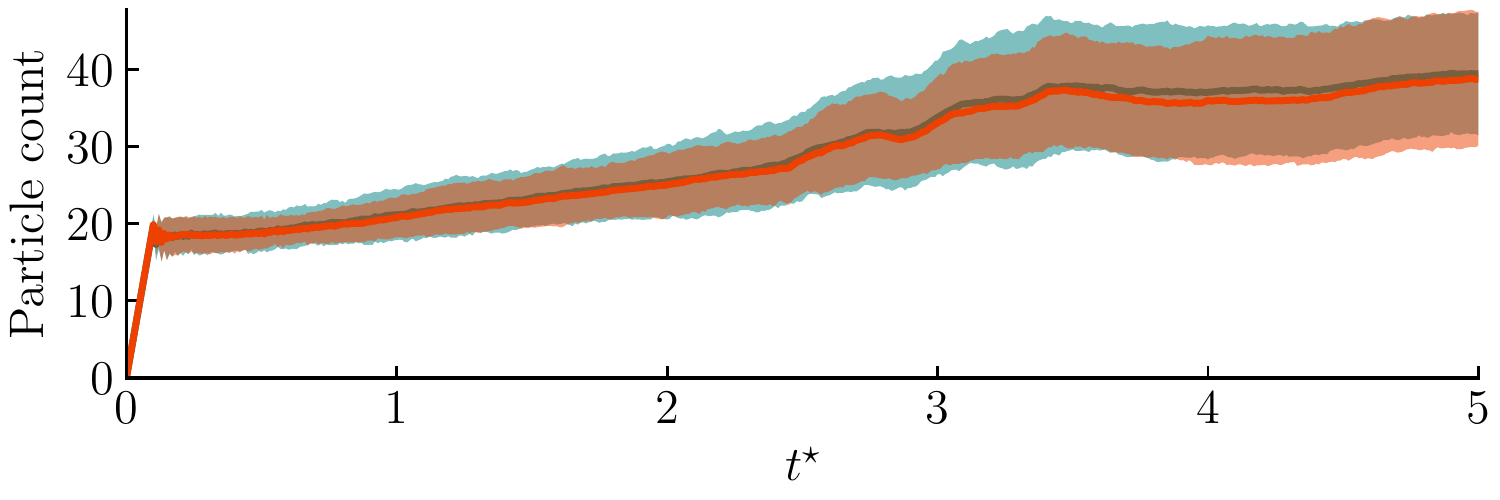}
    \caption{Time history of the particle count for an impulsively translating plate at $20^\circ$ subject to actuation,  averaged over $100$ realizations, from the inviscid vortex model with \senkf{} \lcolor{cenkf} and the inviscid vortex model with \etkf{} \lcolor{cetkf}. Shaded areas show the $95\%$ confidence interval for  the inviscid vortex model with \senkf{} and \etkf{}.}
    \label{fig:case4_count}
\end{figure}

\begin{figure}[tbhp]
    \centering
    \includegraphics[width = \textwidth]{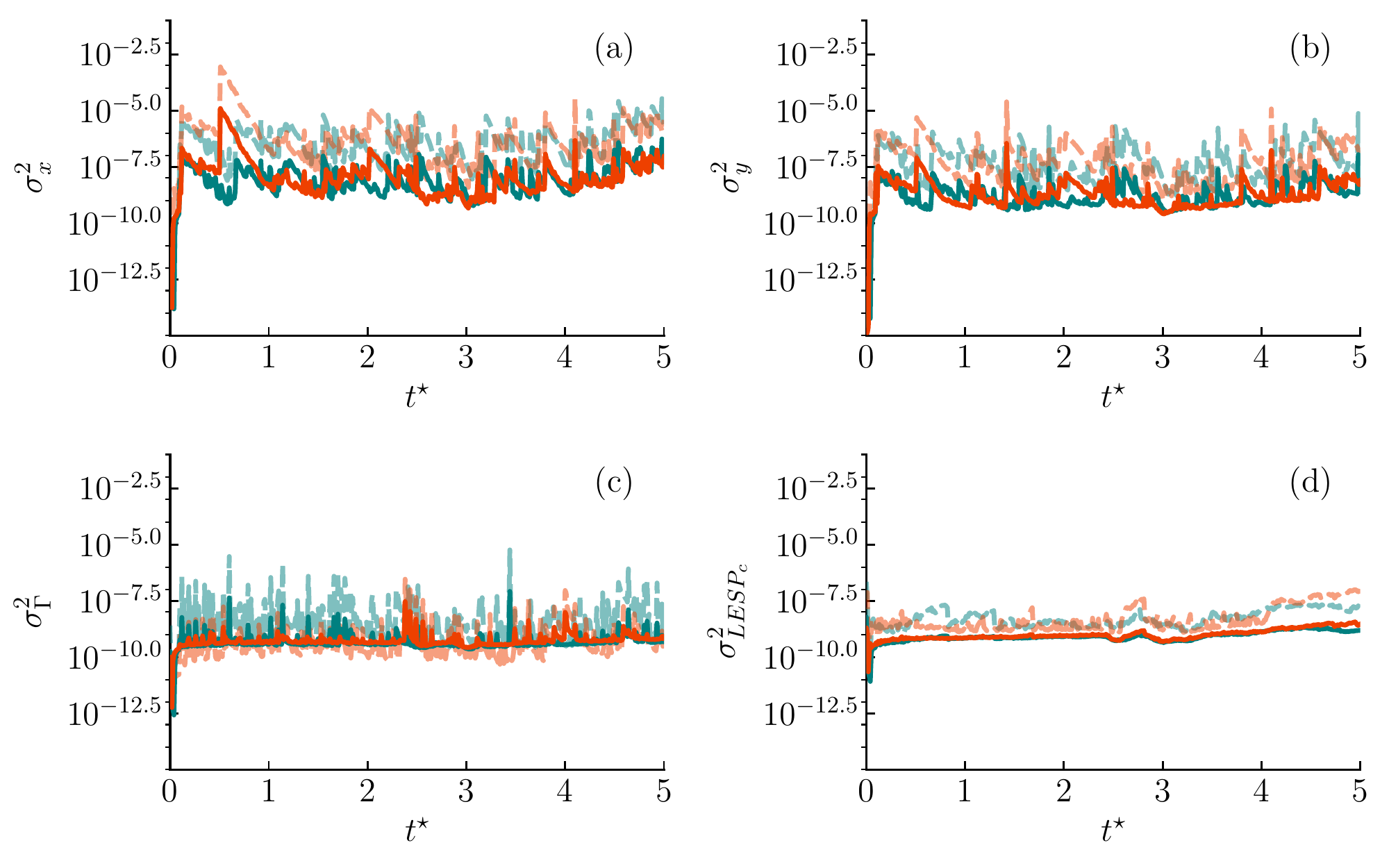}
    \caption{Time history of the ensemble variances of an impulsively translating plate at $20^\circ$ subject to actuation, averaged over $100$ realizations. Mean variances for (a) the $x$ coordinate of the blobs, (b) the $y$ coordinate of the blobs, (c) the circulation of the blobs, and (d) the \lespc{} from the inviscid vortex model with \senkf{} \lcolor{cenkf} and the inviscid vortex model with \etkf{} \lcolor{cetkf}. Shaded areas show the $95\%$ confidence interval for the inviscid vortex model with \senkf{} and \etkf{}.}
    \label{fig:case4_cov}
\end{figure}

The population histories of vortex elements is depicted in Fig.~\ref{fig:case4_count}. The histories are nearly identical for each filter. In the initial time steps, the number of vortex elements increases linearly from $0$ to $20$; the model prevents aggregation of elements during this interval. Subsequently, the population grows slowly up to $35 \pm 7$ at $\ts = 5$. It should be noted that, without aggregation, this population would be approximately 1000 vortex elements (500 time steps, with two elements shed per step). The variation in population among the different realizations of each filter is attributable to the variation in \lespc{}, which sets the initial strengths of the elements, which in turn affects their later aggregation. It is interesting to note that this variation of vortex element populations is proportionally larger than the variation in the pressure and normal force, indicating that there is some non-uniqueness in the mapping from surface pressures to vortex element dynamics.


Fig.~\ref{fig:case4_cov} depicts the ensemble variances of the positions and strengths of the vortex elements and the \lespc{} estimate, averaged over the 100 realizations of each filter. Each variance is lower-bounded by the additive covariance inflation to avoid filter divergence. The variances of the strengths of the vortices and \lespc{} are fairly constant (and near the values set by the inflation parameter) while those of the $x$ and $y$ positions are more variable. These values are larger than the variance set by the inflation, likely due to additional error incurred by aggregation.


\subsection{\label{sub:plate20cylinder}Plate in the wake of a cylinder}

\begin{figure}[tbp]
    \centering
    \includegraphics[width = 0.7\linewidth]{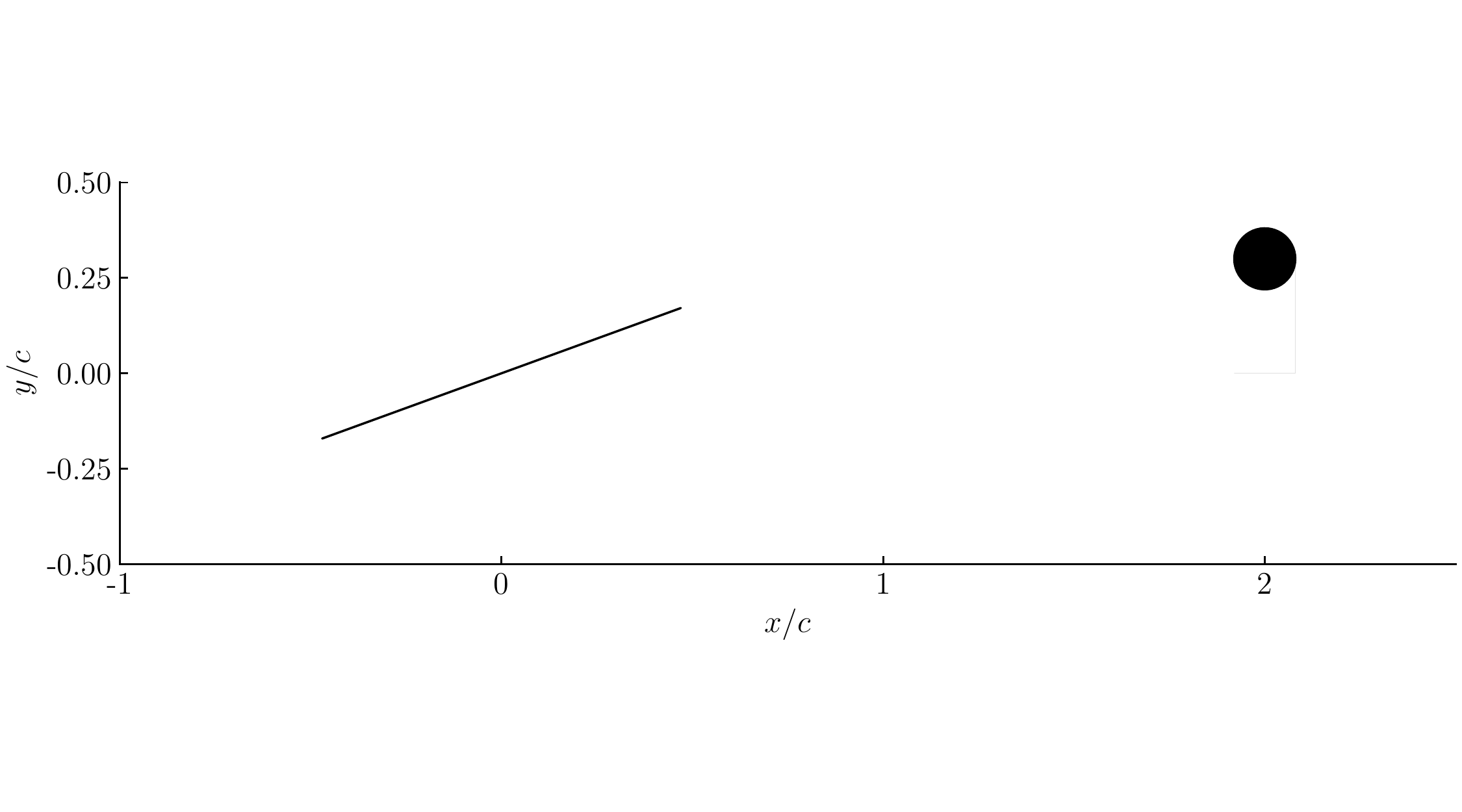}
    \caption{Schematic of a plate at $20^\circ$ behind a cylinder. The plate and cylinder translate to the right at uniform speed.}
    \label{fig:cylinder}
\end{figure}

In this part, we assess our flow estimator when applied to a plate at $20^\circ$ angle of attack in the wake of a cylinder. In the truth system, high fidelity simulations are conducted of a cylinder of diameter $0.16c$ is centered $2$ chord lengths upstream and $0.3$ chord lengths above the plate's centroid, as shown in Fig.~\ref{fig:cylinder}. The plate and cylinder are both impulsively set in motion at $\ts=0$ at speed $U$, so that their relative configuration remains fixed for all time. The Reynolds number based on the cylinder diameter is $80$, sufficiently large that the cylinder's wake exhibits a von K\'arm\'an vortex street. The presence of the plate triggers the wake to break symmetry and achieve this vortex street. This flow configuration distills the main features of a vehicle flying through the wake of a structure, e.g., buildings in an urban environment or other flying vehicles.


\begin{figure}[tbh]
    \centering
    \includegraphics[width  = 1.0\textwidth]{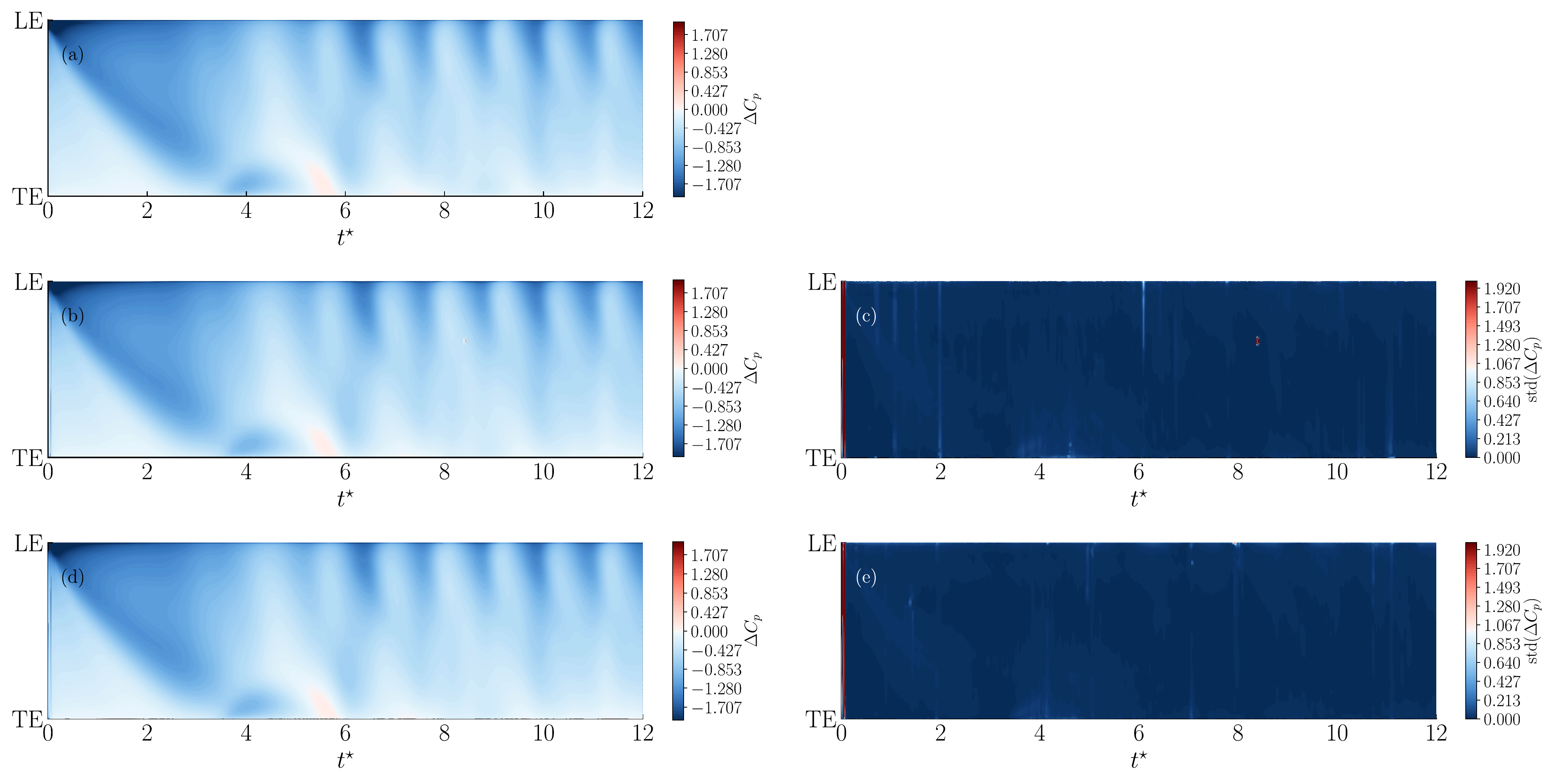}
    \caption{Left column [(a), (b), (d)]: Spatiotemporal map of the pressure coefficient jump for an impulsively translating plate at $20^\circ$ in a cylinder wake (a) high-fidelity numerical simulation at Reynolds number $500$, mean over $100$ realizations of an inviscid vortex model with the \senkf{} (b) and  the \etkf{} (d). Right column [(c), (e)]: Spatiotemporal map of standard deviation of the pressure coefficient jump  over $100$ realizations for the \senkf{}(c) and \etkf{} (e)}
    \label{fig:cylinder_pressure}
\end{figure}

\begin{figure}[tbhp]
\centering
\includegraphics[width =\linewidth]{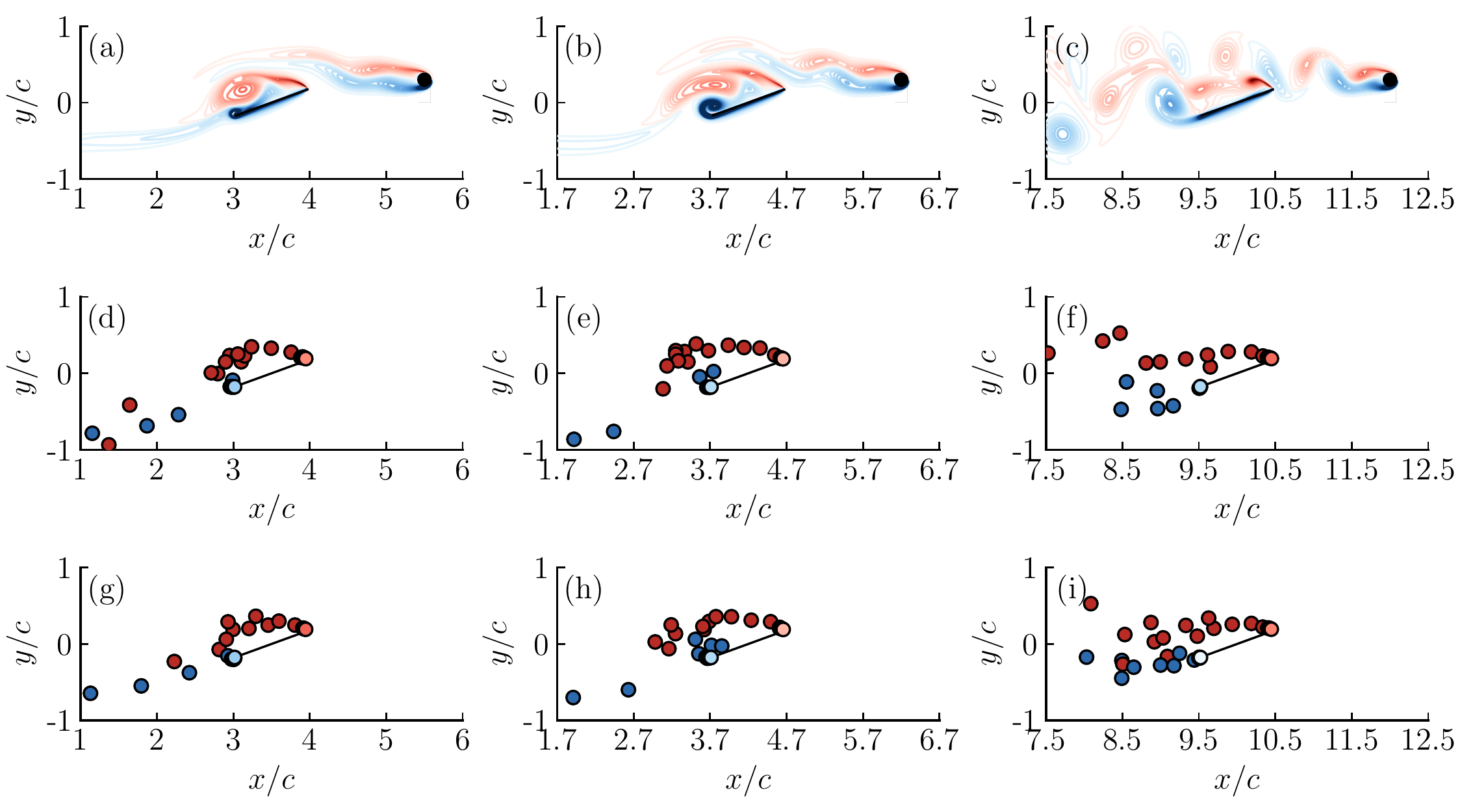}
\caption{Snapshots of the vorticity distribution at $\ts = 3.5$ (left column), $\ts = 4.2$ (middle column) and $\ts = 10.0$ (right column) for an impulsively translating plate at $20^\circ$ in a cylinder wake, predicted from [(a)-(c)] high-fidelity numerical simulation at Reynolds number $500$, [(d)-(f)] inviscid vortex model with \senkf{}, and [(g)-(i)] inviscid vortex model with \etkf{}.}
\label{fig:cylinder_vorticity}
\end{figure}

\begin{figure}[tbhp]
    \centering
    \includegraphics[width = 0.6\textwidth]{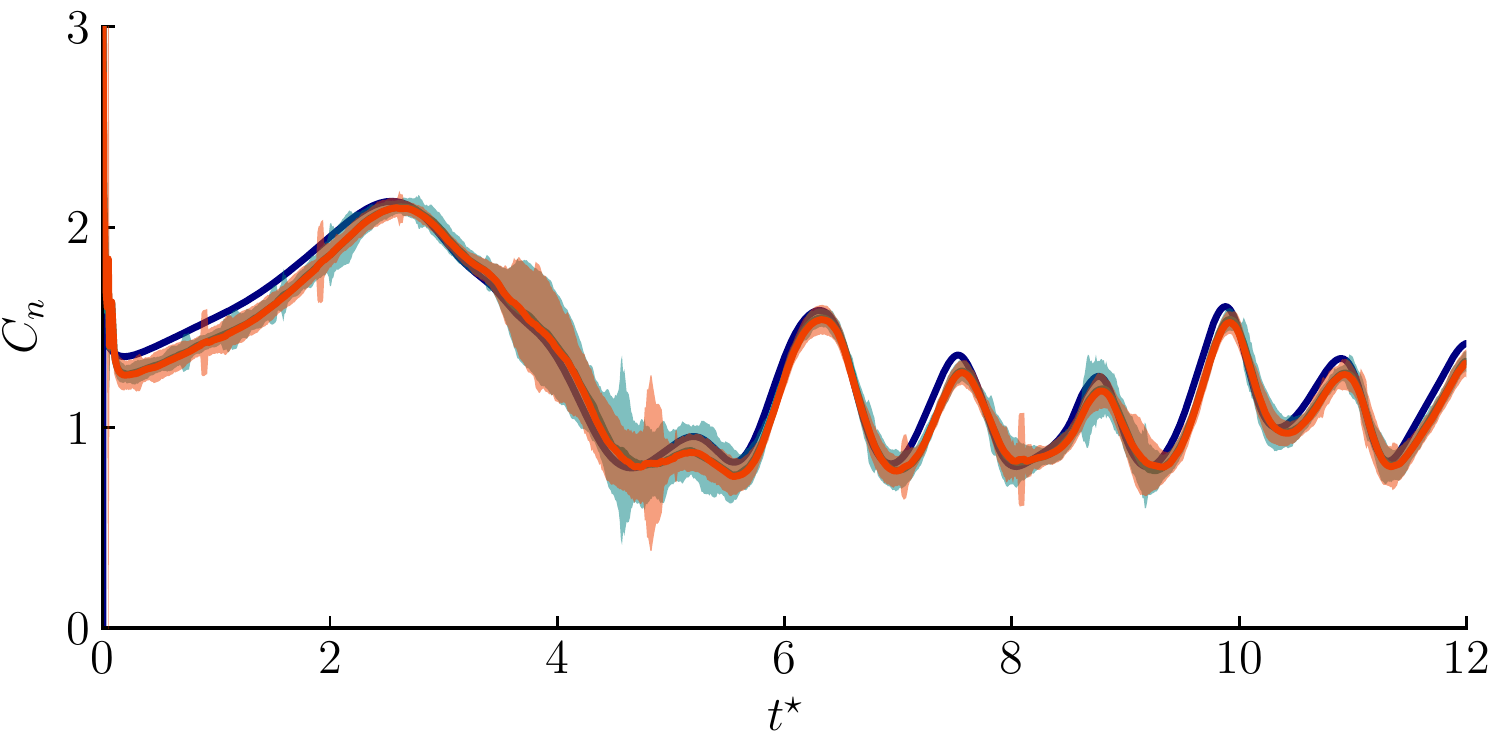}
    \caption{Normal force coefficient of an impulsively translating plate at $20^\circ$ in a cylinder wake, from high-fidelity numerical simulation at Reynolds number $500$  \lcolor{ccfd}, mean over $100$ realizations of the inviscid vortex model with \senkf{} \lcolor{cenkf}, mean over $100$ realizations of the inviscid vortex model with  \etkf{} \lcolor{cetkf}. Shaded areas show the $95\%$ confidence interval for the inviscid vortex model with \senkf{} and \etkf{}}
    \label{fig:cylinder_force}
\end{figure}

\begin{figure}[tbhp]
    \centering
   \includegraphics[width = 0.6\textwidth]{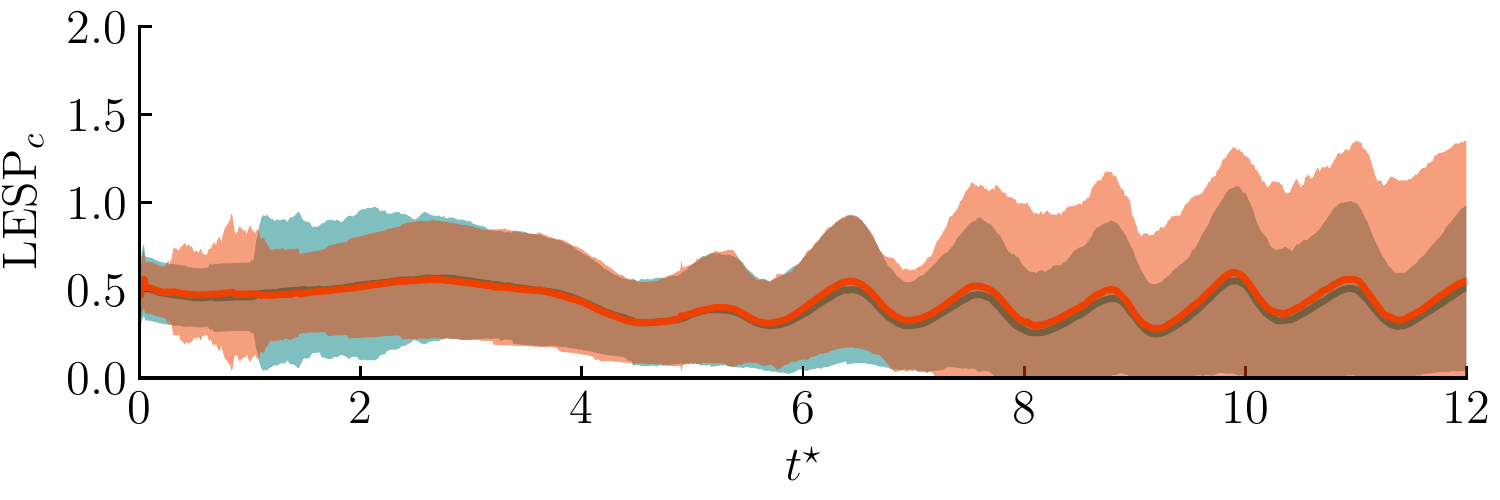}
    \caption{Time history of the ensemble mean value of the \lespc{} of an impulsively translating plate at $20^\circ$ in a cylinder wake, averaged over $100$ realizations, from the inviscid vortex model with \senkf{} \lcolor{cenkf} and the inviscid vortex model with \etkf{} \lcolor{cetkf}. Shaded areas show the $95\%$ confidence interval for the inviscid vortex model with \senkf{} and \etkf{}.}
    \label{fig:cylinder_lespc}
\end{figure}

\begin{figure}[tbhp]
    \centering
    \includegraphics[width = 0.6\linewidth]{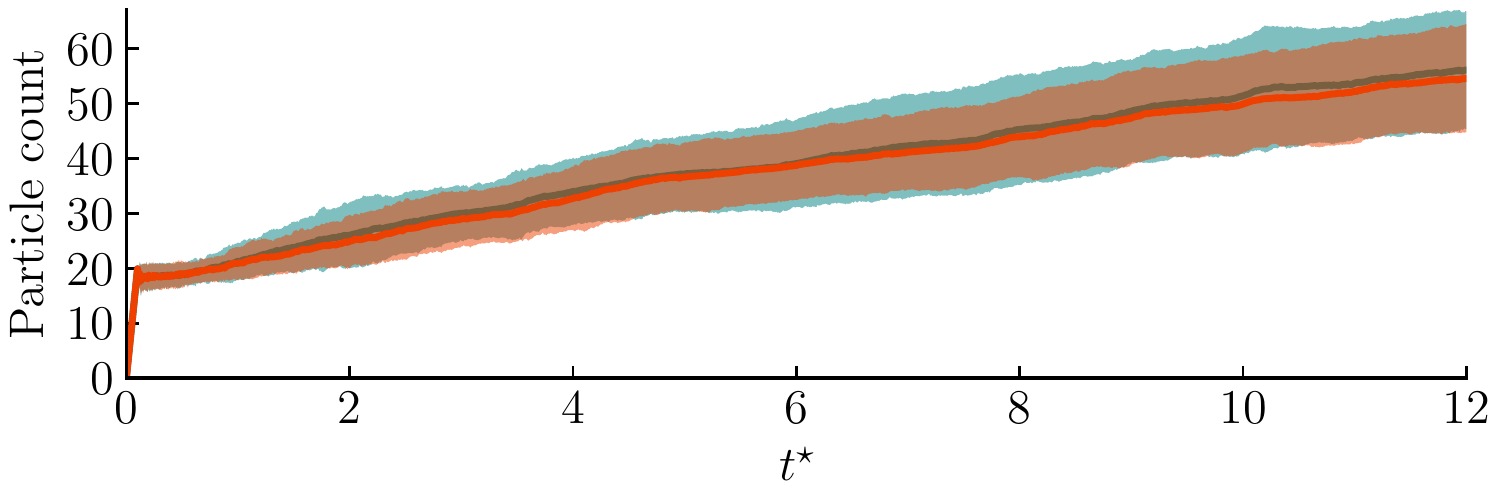}
    \caption{Time history of the particle count for an impulsively translating plate at $20^\circ$ in a cylinder wake,  averaged over $100$ realizations, from the inviscid vortex model with \senkf{} \lcolor{cenkf} and the inviscid vortex model with \etkf{} \lcolor{cetkf}. Shaded areas show the $95\%$ confidence interval for  the inviscid vortex model with \senkf{} and \etkf{}.}
    \label{fig:cylinder_count}
\end{figure}

It is important to recall that we do not include the cylinder and its wake in our vortex model flow estimator; its effects are only felt through the pressure jump measurements obtained from surface sensors along the plate. Fig.~\ref{fig:cylinder_pressure} depicts the history of the true surface pressure distribution and those estimated by the \senkf{} and \etkf{}; the associated standard deviations are shown on the right. Both filters estimate pressure fields that agree very well with the true distribution. As expected, it takes about $2$ convective times for the cylinder wake to reach and be sensed by the plate. Over this time window, the pressure field is essentially disturbance-free. The encounter of the vortex structures shed by the cylinder leaves successive short-lived pressure disturbances.

The true vorticity distribution and mean sets of vortex elements are compared in Fig.~\ref{fig:cylinder_vorticity} at $\ts = 3.5$ (soon after the wake has reached the plate), at $\ts = 4.2$ (when the cylinder wake transitions to a von K\'arm\'an vortex street) and at $\ts = 10$ (long after the flow has achieved periodic vortex shedding). As expected, the vortex models do not attempt to represent the cylinder wake with vortex elements; rather, the filter accommodates the influence of the wake vorticity by modifying the behavior of the vortex elements shed from the plate. Large scale structures of the flow around the plate are captured and match visually with the true vorticity field.

Concomitantly, the normal force estimate agrees well with the truth, as seen in Fig.~\ref{fig:cylinder_force}. The transition from a symmetric cylinder wake to a periodic vortex shedding causes some challenge to the discrete vortex models, as highlighted by the temporary growth of the uncertainty envelope around $\ts = 4.0$. The envelope remains small for both filters after the periodic wake behavior has been established. The history of the \lespc{} estimate in Fig.~\ref{fig:cylinder_lespc} reveals the same growth and decay of this critical value observed in the previous example; here, they represent a response to individual cylinder wake vortices passing the leading edge of the plate. These variations are essential to our inviscid framework to control the leading-edge vortex shedding in the presence of flow perturbations. As in the previous example, the weak correlation between the \lespc{} and measured pressure causes significant volatility in the estimated \lespc{} among the different realizations. The vortex element population remains small, $O(60)$, after 12 convective time units. Without aggregation, we would have to track $2400$ vortex elements.



\subsection{Dissection of the EnKF analysis}

The results shown thus far depict the overall evolution of the state with time, but do not reveal the manner in which individual components of the state---vortex element positions and strengths, and \lespc{}---are modified by the analysis step to account for pressure mismatch. Here, we examine this more closely, and particularly, ask whether there is some notion of \textit{locality} inherent in the radius of action of a given pressure measurement on the correction of the state components. This question motivates our dissection of the input-output structure of the Kalman gain matrix $\K \in \real^{n,d}$ (\ref{eqn:kalman gain}), the linear map from the $d$ sensors (measurement innovation) to the $n$-dimensional state correction. The component $K_{ij}$ of the Kalman gain tells us how much the $j$th measurement updates the $i$th component of the state; a low value (in magnitude) indicates a low impact. As we recall from (\ref{eqn:statea}), the Kalman gain is derived from the ensemble's forecast anomalies.



\begin{figure}[tbp]
    \centering
    \includegraphics[width = \linewidth]{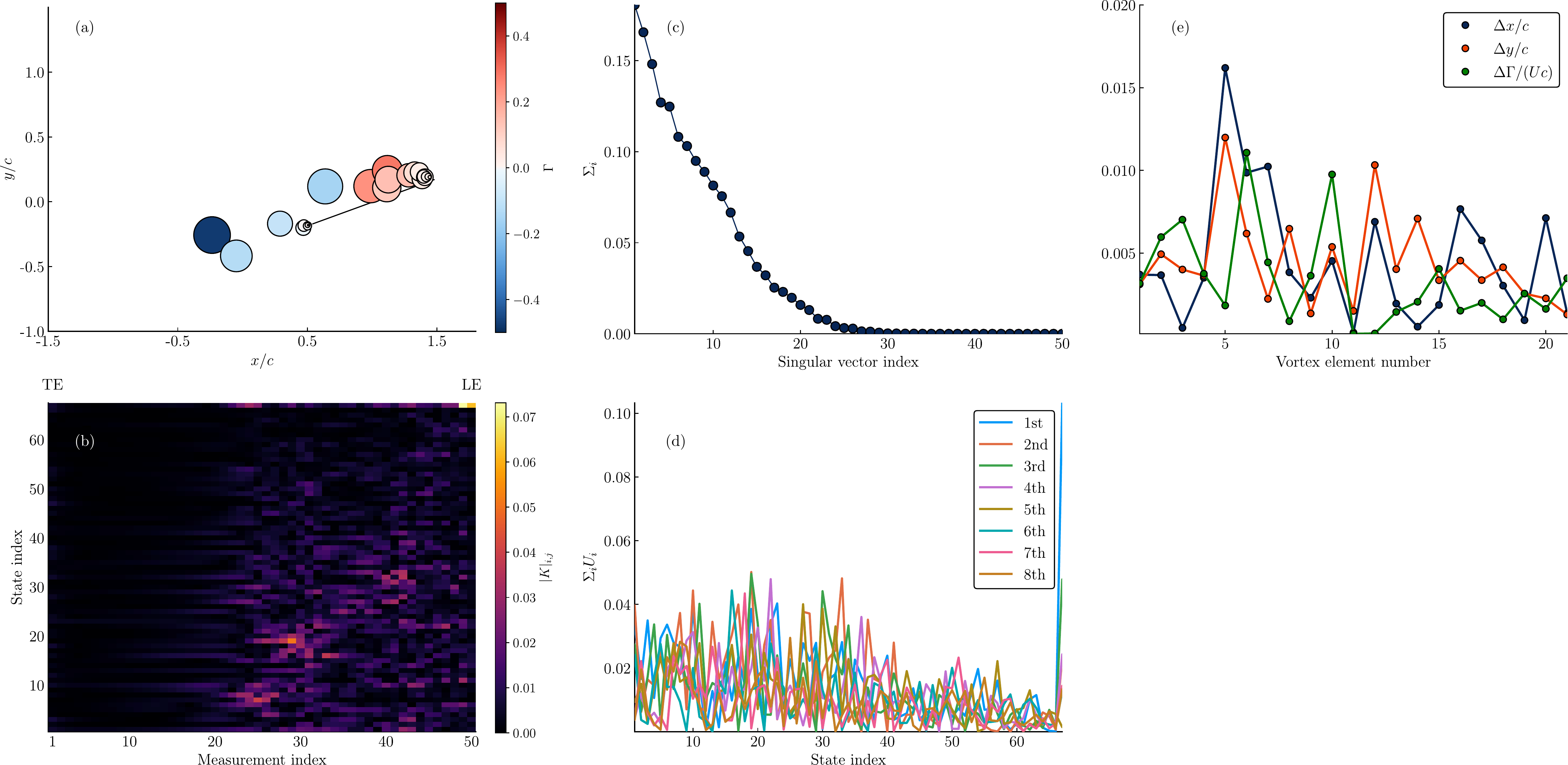}
    \caption{Study of the analysis step at $\ts = 1$ for the impulsively translating plate in the cylinder wake.
    (a): Vorticity distribution from the discrete vortex model with ETKF. Sizes indicate  "age" of the vortices (large means old). Colors indicate vorticity sign (blue is negative). (b): Element-wise magnitude of the Kalman gain. (c): Singular values of the Kalman gain. (d): First eight left singular vectors of the Kalman gain scaled by their associated singular value. (e): Mean difference of $x$, $y$ position and circulation of vortices during the analysis step.}
    \label{fig:cylinder_K1}
\end{figure}

\begin{figure}[tbp]
    \centering
    \includegraphics[width = \linewidth]{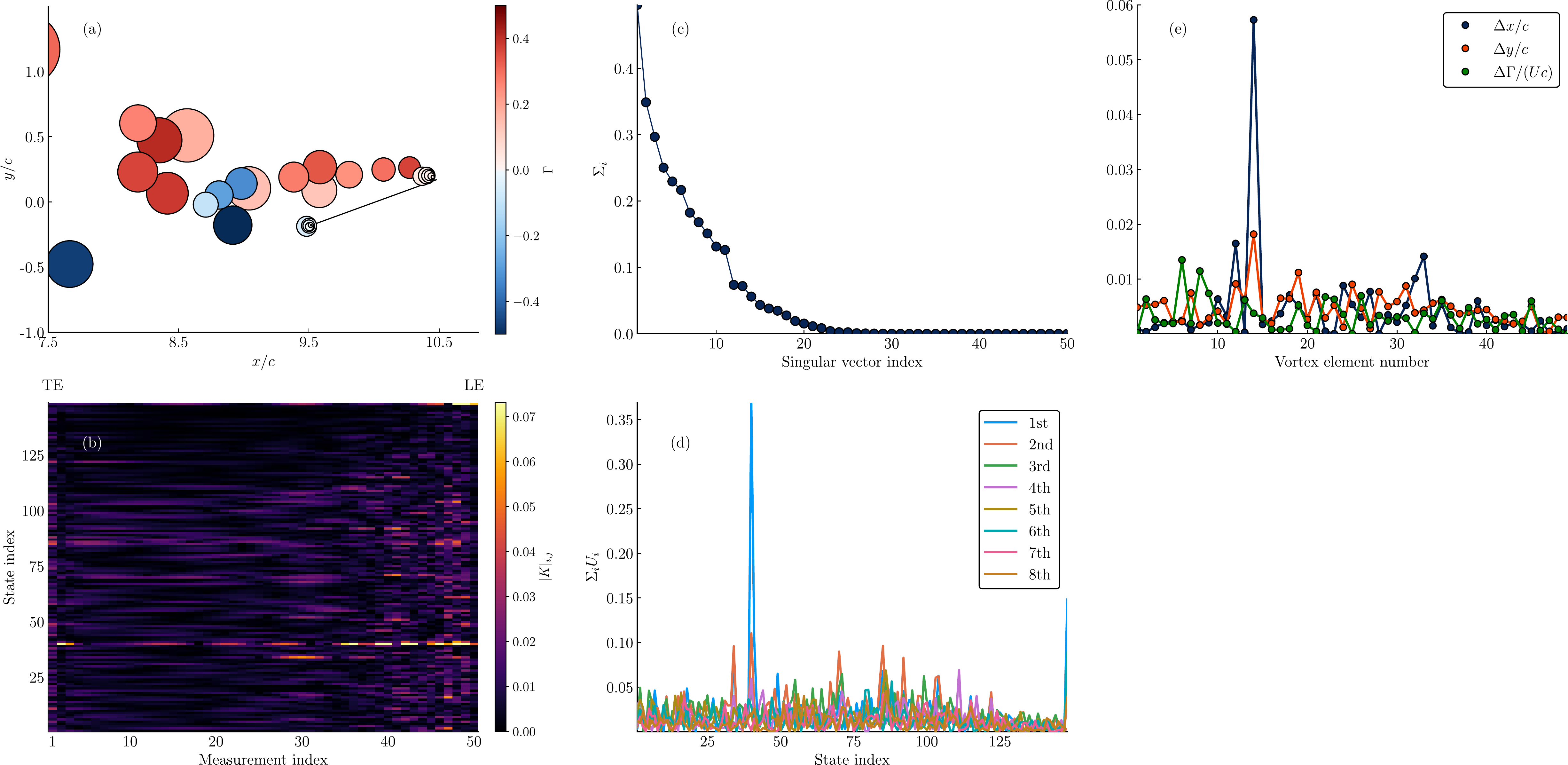}
    \caption{Study of the analysis step at $\ts = 10$ for the impulsively translating plate in the cylinder wake.
    (a): Vorticity distribution from the discrete vortex model with ETKF. Sizes indicate  "age" of the vortices (large means old). Colors indicate vorticity sign (blue is negative). (b): Element-wise magnitude of the Kalman gain. (c): Singular values of the Kalman gain. (d): First eight left singular vectors of the Kalman gain scaled by their associated singular value. (e): Mean difference of $x$, $y$ position and circulation of vortices during the analysis step.}
    \label{fig:cylinder_K10}
\end{figure}

Figs.~\ref{fig:cylinder_K1}, \ref{fig:cylinder_K10} examine how the \etkf{} algorithm modifies the state during the analysis step at $\ts = 1$ and $10$, respectively. We use a different visualization convention for the vortex element distribution in Fig.~\ref{fig:cylinder_K1}(a) and \ref{fig:cylinder_K10}(a). The size of each element inversely indicates its index in the collection of vortices. Since newly-added elements are assigned indices at the end of the list, larger circles correspond to ``older'' vortices. Colors still indicate vorticity sign and strength. With this convention, it is easier to see which vortex elements are updated by the filter. At $\ts = 1$, the state vector has dimension $n = 67$, corresponding to $22$ vortex elements and the \lespc{} estimate. Fig.~\ref{fig:cylinder_K1}(b) shows a heat map of the magnitudes of the Kalman gain elements. Low measurement index corresponds to a pressure sensor near the trailing edge. The \lespc{} estimate is always the last component of the state vector. Overall, the magnitudes of the components of the Kalman gain are small. The effect of these small gain values is highlighted in Fig. \ref{fig:cylinder_K1}(e), which shows the mean analysis updates to the non-dimensional $x$ and $y$ positions and circulations of vortex elements. These updates are comparable to or smaller than the time step size $0.01$, and thus, of similar order of magnitude to their forecast updates.



It is useful to recall the relationship between pressure and vorticity in incompressible flow. Though we formally compute the pressure in the vortex model with the unsteady Bernoulli equation, it can also be obtained through the following Poisson equation:
\begin{equation}
    \label{eqn:poisson_p}
    \nabla^2 \left(p + \frac{1}{2} \rho |\BB{u}|^2 \right) = \rho \nabla \cdot(\BB{u} \times \BB{\omega}),
\end{equation}
where $\BB{u}$ denotes the velocity field. This velocity depends in part on the contributions from all other vortices (along with the body's motion and/or uniform flow). Thus, this equation clearly illustrates that pressure depends non-linearly on the vorticity field. But even if the model was made linearly dependent on vorticity, e.g., by linearizing about some base flow, the pressure remains non-linearly dependent on the positions of the vortices, due to the ellipticity of relationship (\ref{eqn:poisson_p}). Therefore, the linear analysis step in the Kalman gain might be inadequate to estimate the true modification of a vortex due to a pressure mismatch. Also, the EnKF is a sequential filter, which means that it performs iterated state updates based on pressure difference. But even if each analysis step only slightly modifies the state, the long-term impact of these linear updates is important.

Some interesting physical features can be inferred by a closer study of Fig.~\ref{fig:cylinder_K1}(b). The top line of the heat map shows the impact of pressure measurements on the \lespc{} estimate. Only a few pressure sensors near the leading edge have a high impact on the \lespc{}. This is not surprising, since \lesp{} a measure of the integrated pressure about the leading edge, and \lespc{} is a threshold on this value. The heat map also shows that pressure sensors near the trailing edge do not impact the state substantially. For a given vortex element, the most impactful sensors are those that are closest. Most of the state update are due to sensors between the leading edge and mid-chord, consistent with the position of the leading-edge vortex about the mid-chord at $\ts = 1$. The newly released vortex elements are updated by the pressure sensors in the immediate vicinity of the leading-edge.

To explore the update process more deeply, we perform a singular value decomposition of the Kalman gain matrix:
\begin{equation}
    \K = \BB{U} \BB{\Sigma} \BB{V}^\top,
\end{equation}
where $\K \in \real^{n,d}$, $\BB{U} \in \real^{n,d}$, $\BB{\Sigma} \in \real^{d,d}$, $\BB{V} \in \real^{d,d}$. The columns of $\BB{U}$ and $\BB{V}$ are called the left and right singular vectors, and can also be interpreted here as modes of the response (of the state) and forcing (from the measurements), respectively. The response-forcing pairs are ranked by decreasing gain, given by the diagonal entries of the diagonal matrix $\BB{\Sigma}$, the singular values; these are shown in Fig.~\ref{fig:cylinder_K1}(c) for the Kalman gain at $\ts=1$. Each measurement mode is transformed into a response mode, amplified by their corresponding singular value. The decay rate of the singular values is a useful indication of the effective number of pressure measurements used in the analysis step. Here, fourteen modes are necessary to capture $99\%$ of the energy (sum of the square of the singular values). The response modes tell us which state indices are the most impacted.  Fig.~\ref{fig:cylinder_K1}(d) depicts the first eight response modes that account for $90\%$ percent of the energy. The largest updates are applied to the \lespc{} estimate (the largest state index) and to the oldest vortices with large circulation. Though the results in Fig.~\ref{fig:cylinder_K1} correspond to the \etkf{}, we have also found similar results for the \senkf{}. 

Fig.~\ref{fig:cylinder_K10} shows equivalent plots for $\ts = 10$. The update of the \lespc{} is still driven by the sensors near the leading edge and a few about the mid-chord. The sensors near the leading edge still have a stronger impact on the state correction. The SVD of the Kalman gain reveals in Fig.~\ref{fig:cylinder_K10}(d) that the \lespc{} and a particular vortex element are the most affected components of the state. From Fig. \ref{fig:cylinder_K10}(e), this particular element's update is displaced by $0.06c$, $0.02c$ in the $x$  and $y$ directions, respectively. The element is located five chord lengths away from the plate, an unlikely recipient of update from any surface measurements, particularly to sensors near the leading edge as panel (b) indicates. Similarly, we do not expect that pressure sensors about the mid-chord will update the \lespc{} value. These spurious updates reveal a flaw of using a finite ensemble size.

In future work, we will explore the concept of \textit{localization} to remove these artificial correlations by performing a local state update \citep{houtekamer1998data, Evensen2003TheImplementation, asch2016data}. In other words, pressure measurements will be assimilated one at a time and only affect state components that are within a certain distance from that specific pressure sensor location. Intuitively, we want to embed into our estimation framework that vortices in the far wake of the plate should not be modified by pressure discrepancies on the plate. Localization is another form of regularization, complementary to the covariance inflation that helps overcome the rank deficiency of the covariance matrices. By performing local analyses, each reduced inverse problem can be made full rank and improves the overall performance of the filter \cite{asch2016data}.

\section{\label{sec:conclusion}Conclusions}

This paper presents several refinements of the framework introduced by Darakananda et al. \cite{darakananda2018data} for the estimation of two-dimensional unsteady aerodynamic flows. As in \cite{darakananda2018data}, the dynamical model is a discrete vortex model that advects a collection of regularized vortices, successively released from the edges of an infinitely thin plate and aggregated to maintain a modest population. The unrepresented physical features of the vortex model, e.g., effects of viscosity, are accounted for by the assimilation of pressure measurements obtained from the true physical process. Vorticity flux from the leading edge is obtained by a critical leading-edge suction parameter whose value is also obtained from the measurements.

The data assimilation has been carried out with the ensemble Kalman filter (EnKF), which relies on a finite-sized ensemble of vortex models to forecast the state and uncertainty of the system. We have interpreted this ensemble forecast physically as an effective translation, stretching, and rotation of the elliptically shaped regions defining each vortex element's uncertainty region. We have explored two different versions of the EnKF: the stochastic ensemble Kalman filter (\senkf{}), which was also used in \cite{darakananda2018data}, and the ensemble transform Kalman filter (\etkf{}). The \senkf{} introduces additional numerical noise by not exactly obtaining the correct covariance propagation during the analysis step. In the current vortex estimation procedure, this noise manifests itself as spurious pressures exerted on the surface of the wing. The \etkf{} \cite{bishop2001adaptive}, in contrast, reproduces the exact covariance relation even for finite sized ensembles. We have found that this latter filter greatly mitigates the appearance of spurious pressures on the surface, and clusters of vortex elements match better with the true coherent structures. As in \cite{darakananda2018data}, multiplicative and additive covariance inflation are used to prevent filter divergence. Overall, we have found that the \etkf{} has better performance and robustness than the \senkf{}.


Two flow configurations have been considered to assess the performance of our flow estimator and the influence of the filtering algorithm. The true pressure measurements were obtained from high-fidelity numerical simulations of an impulsively translating plate at $20^\circ$ and Reynolds number $Re = 500$, subject to flow disturbances. The first example examined the response to a sequence of strong body force pulses applied near the leading edge mimicking flow actuation. We demonstrate that the ensemble of data-assimilated vortex models accurately capture the non-linear flow interactions between the pulses' responses. It should be stressed that a linear convolutional model, with a convolution kernel based on a single pulse, would fail due to these non-linear interactions. The second example consisted of the response of a plate subject to the large scale and coherent perturbations created in the wake of an upstream cylinder. The normal force coefficient on the plate was accurately estimated over 12 convective time units. It should be stressed that the estimation framework does not require a model of the disturbance itself. The disturbance's effect on shed vorticity is accounted for by using measurements to update the state vector in the filter analysis step. This feature is particularly beneficial for a real-time aerodynamic flow estimator, in which the structure of incident flow disturbances such as gusts are unknown.


It is important to stress that our framework is independent of the source of the measurements, and our objective in ongoing work is to assimilate  measurements from a small number of surface pressure transducers in a physical experiment or on board a flight vehicle. However, there are still challenges to address first before we can achieve this. In particular, we are exploring ways to further improve the analysis step. In this work, with the help of an SVD of the Kalman gain, we have illuminated the function of this step by examining how pressure discrepancies modify the state components. This examination has shown that, in most cases, the Kalman gain ensures that sensors primarily affect state components that are physically close, e.g., vortex elements near the plate or the value of \lespc{}. However, the finite ensemble size can lead to spurious long-range correlations and non-physical updates of the vortices in the far wake. To mitigate this issue, we are currently investigating localization, which aims to remove these long-range correlations by performing local analyses. We are also exploring approaches to systematically sparsify the sensors needed for analysis updates.


\section{acknowledgments}
Support by the U.S. Air Force Office of Scientific Research FA9550-18-1-0440 is gratefully acknowledged. The authors also greatly benefited from discussions with Tim Colonius. 

\appendix

\section{\label{apx:enkf}Ensemble Kalman filter algorithms}

\begin{algorithm}[tbhp]
 Initialize $M$ ensemble member $(\state^{a,i}_0) \sim \N(\statebar_0, \BB{P}_0)$ \\
 \For{$k= 1:K$}{
 \quad \\
 \quad Compute the ensemble forecast:\\
 \quad \quad $\state^{f,i}_k = \dyn(\state^{a,i}_{k-1}) + \sampledyn_{k-1}^i$, for $i=1, \ldots, M$\\
 \quad Sample from the measurement noise distribution $\pi_{\noiseobs_k}$:\\
 \quad \quad $\sampleobs_k^i \sim \N(\zero, \covobs_k)$, for $i=1, \ldots, M$\\
 \quad Compute the sample means:\\
 \quad \quad $\statebarf_k = \frac{1}{M} \sum_{i=1}^{M} \state^{f,i}_k$, \quad $\overline{\sampleobs}_k = \frac{1}{M} \sum_{i=1}^{M} \sampleobs_k^i$, \quad
 $\measbarf_k =\frac{1}{M} \sum_{i=1}^{M} \obs(\state^{f,i}_k)$\\
 \quad and the anomalies:\\
 \quad \quad $\X^{\prime f,i}_k = (\state^{f,i}_k - \statebarf_k)/\sqrt{M-1}$,\quad  $\Ypfi_k = (\obs(\state^{f,i}_k) - \sampleobs^i_k -\measbarf_k + \overline{\sampleobs}_k)/\sqrt{M-1}, \mbox{ for } i=1, \ldots, M$\\
 \quad Invert these equations for $(\BB{b}^i_k)$:\\
 \quad \quad $\Ypf \Ypf^\top \BB{b}^i_k = \meas^\star_k + \sampleobs_k^i - \obs(\state^{f,i}_k), \mbox{ for } i=1, \ldots, M$\\
 \quad Update the ensemble:\\
 \quad \quad $\state^{a,i}_k = \state^{f,i}_k + \X_k^{\prime, f} {\BB{Y}_k^{\prime, f}}^{\top}\BB{b}^i_k$
 }
\caption{Algorithm for the \senkf{}, adapted from Asch et al.  \cite{asch2016data}} \label{algo:senkf}
\end{algorithm}

\begin{algorithm}[tbhp]
 Input: integer $M$\\
 Draw $(M-1)\times(M-1)$ samples from an standard normal distribution and store them in $\BB{\Omega} \in \real^{(M-1)\times (M-1)}$:\\
 \quad $\BB{\Omega} = \mbox{randn}(M-1,M-1)$\\
 SVD decomposition of $\BB{\Omega}$:\\
 \quad $\BB{\Omega} = \U \BB{\Sigma} \BB{V}^\top$\\
 \quad $\BB{\Omega} = \BB{V}^\top$\\
 Compute $\BB{b} = \one/\sqrt{M}$\\
 Compute $\BB{b}^{sign}$\\
 \quad $\BB{b}^{sign} = \BB{b}$\\
 \quad $\BB{b}^{sign}[M] \mathrel{+}= \mbox{sign}(\BB{b}[M])$ \# add $\mbox{sign}(\BB{b}[M])$ to the last entry\\
 Initialize $\BB{B} \in \real^{M\times M}$\\
 Fill columns of $\BB{B}$:\\
 \quad $\BB{B}[:,1] = \BB{b}$\\
 \quad $\BB{B}[:,2:M] = -(\id_M + \BB{b}^{sign}{\BB{b}^{sign}}^\top/(||\BB{b}|| + 1))[:,1:M-1]$\\
 Initialize $\BB{\Lambda_b} \in \real^{M\times M}$\\
 Fill $\BB{\Lambda_b}$\\
 \quad $\BB{\Lambda_b}[1,1] = 1$\\
 \quad $\BB{\Lambda_b}[1:M-1, 1:M-1] = \BB{\Omega}$\\
 return $\BB{U} = \BB{B}\BB{\Lambda_b}\BB{B}^\top$
\caption{Algorithm to generate mean-preserving random rotations, adapted from \citep{todter2015second, nerger2012unification}}
\label{algo:rand}
\end{algorithm}

\begin{algorithm}[tbhp]
$\U$ is a mean-preserving rotation generated from time to time, see Algorithm~\ref{algo:rand}, otherwise set to $\id_M$\\ 
 Initialize $M$ ensemble member $(\state^{a,i}_0) \sim \N(\state_0, \BB{P}_0)$ \\
 \For{$k= 1:K$}{
 \quad \\
 \quad Compute the ensemble forecast:\\
 \quad \quad $\state^{f,i}_k = \dyn(\state^{a,i}_{k-1}) + \sampledyn_{k-1}^i$ for $i=1, \ldots, M$\\
 \quad Compute the prior mean and anomaly matrix:\\
 \quad \quad $\statebarf_k = \frac{1}{M} \sum_{i=1}^{M} \state^{f,i}_k$, \quad $\X^{\prime f,i}_k = (\state^{f,i}_k - \statebarf_k)/\sqrt{M-1}, \mbox{ for } i=1, \ldots, M$\\
 \quad Compute the observations:\\
 \quad \quad $\BB{Y}_k^{f,i} = \obs(\state^{f,i}_k)$,  $\mbox{ for } i=1, \ldots, M$\\
 \quad Compute mean of the observations $\measbarf_k$:\\
 \quad \quad $\measbarf_k =\frac{1}{M}\BB{Y}_k^{f} \one$\\
 \quad Construct the matrices $\BB{S}$ and $\BB{G}$:\\
 \quad \quad $\BB{S} = \covobs^{-1/2}_k(\BB{Y}_k^{f} - \measbarf_k \one^\top)/\sqrt{M-1}$\\
 \quad \quad $\G = (\id_M + \BB{S}\BB{S}^\top)^{-1}$\\
  \quad Construct the vectors $\BB{\delta}, \BB{c}$:\\
   \quad \quad $\BB{\delta} =  \covobs^{-1/2}_k(\meas^\star_k  - \measbarf_k)  $\\
 \quad \quad $\BB{c} = \G \BB{S}^\top \BB{\delta}$\\
 \quad Update the ensemble:\\
 \quad \quad $\Xa_k = \statebarf_k \one^\top + \X^{\prime f}_k(\BB{c} \one^\top + \sqrt{M-1}\sqG \U)$
 }
\caption{Algorithm for the \etkf{}, adapted from Asch et al.~\cite{asch2016data}} \label{algo:etkf}
\end{algorithm}


 

\bibliography{mybib}

\bibliographystyle{unsrt}

\end{document}

%% file: macros.tex
\DeclareMathOperator*{\argmin}{argmin}

\newcommand{\BB}{\boldsymbol}
\newcommand{\be}{\begin{equation}}
\newcommand{\ee}{\end{equation}}

\newcommand{\ba}{\begin{align}}
\newcommand{\ea}{\end{align}}

\newcommand{\real}{\mathbb{R}}
\newcommand{\id}{\BB{I}}

\definecolor{forecast}{RGB}{19, 24, 143}
\definecolor{analysis}{RGB}{208, 2, 27}

\newcommand\lcolor[1]{\crule[#1]{2mm}{2mm}}

\definecolor{ccfd}{rgb}{0.0,0.0,0.502}
\definecolor{cenkf}{rgb}{0.0,0.502,0.502}
\definecolor{cetkf}{rgb}{0.933,0.251,0.0}  
\definecolor{cetkfconf}{RGB}{246, 157, 127}
\definecolor{cgust}{rgb}{0.196,0.804,0.196}

\newcommand\crule[3][black]{\textcolor{#1}{\rule{#2}{#3}}}

\newcommand{\lesp}{LESP}
\newcommand{\lespc}{$\mbox{LESPc}$}

\newcommand{\ts}{t^\star}

\newcommand{\statefi}{{\state^{f,i}}}

\newcommand{\stateai}{{\state^{a,i}}}
\newcommand{\statebar}{\overline{\state}}
\newcommand{\statebara}{\statebar^a}
\newcommand{\statebarf}{\statebar^f}

\newcommand{\zero}{\BB{0}}
\newcommand{\one}{\BB{1}}

\newcommand{\K}{\BB{K}}

\newcommand{\Pa}{\BB{P}^a}
\newcommand{\Pf}{\BB{P}^f}

\newcommand{\Pbar}{\overline{\BB{P}}}
\newcommand{\Pbara}{\overline{\BB{P}}^a}
\newcommand{\Pbarf}{\overline{\BB{P}}^f}

\newcommand{\ensemble}[1]{\left\{#1\right\}}

\newcommand{\X}{\BB{X}}
\newcommand{\Xa}{\X^a}

\newcommand{\Xp}{{\BB{X}^{\prime}}}
\newcommand{\Xpa}{{\BB{X}^{\prime a}}}
\newcommand{\Xpf}{{\BB{X}^{\prime f}}}

\newcommand{\Xbar}{\overline{\BB{X}}}

\newcommand{\covobsp}{{\BB{W}^{\prime}}}
\newcommand{\covobspi}{\BB{W}^{\prime, i}}

\newcommand{\Ypf}{{\BB{Y}^{\prime f}}}
\newcommand{\Ypfi}{\BB{Y}^{\prime f, i}}

\newcommand{\U}{\BB{U}}

\newcommand{\G}{\BB{G}}
\newcommand{\sqG}{{\BB{G}^{1/2}}}

\newcommand\given[1][]{\:#1\vert\:}

\newcommand{\N}{\mathcal{N}}

\def\E#1{\mathrm{E}[#1]}
\def\cov#1{\mathrm{Cov}[#1]}

\newcommand{\senkf}{sEnKF}
\newcommand{\etkf}{ETKF}

\newcommand{\state}{\BB{x}}
\newcommand{\meas}{\BB{y}}
\newcommand{\measbarf}{\overline{\meas}^f}
\newcommand{\dyn}{\BB{f}}
\newcommand{\obs}{\BB{h}}
\newcommand{\Obs}{\BB{H}}
\newcommand{\noisedyn}{\BB{\mathsf{W}}}
\newcommand{\noiseobs}{\BB{\mathsf{V}}}
\newcommand{\sampledyn}{\BB{w}}
\newcommand{\sampleobs}{\BB{v}}

\newcommand{\covobs}{\BB{V}}

\newcommand{\Statex}{\BB{\mathsf{X}}}
\newcommand{\Meas}{\BB{\mathsf{Y}}}


%% file: main.bbl
\begin{thebibliography}{10}

\bibitem{dickinson1993unsteady}
Michael~H Dickinson and Karl~G Gotz.
\newblock Unsteady aerodynamic performance of model wings at low reynolds
  numbers.
\newblock {\em Journal of experimental biology}, 174(1):45--64, 1993.

\bibitem{birch2001spanwise}
James~M Birch and Michael~H Dickinson.
\newblock Spanwise flow and the attachment of the leading-edge vortex on insect
  wings.
\newblock {\em Nature}, 412(6848):729--733, 2001.

\bibitem{milano2005uncovering}
Michele Milano and Morteza Gharib.
\newblock Uncovering the physics of flapping flat plates with artificial
  evolution.
\newblock {\em Journal of Fluid Mechanics}, 534:403--409, 2005.

\bibitem{taira2009three}
Kunihiko Taira and TIM Colonius.
\newblock Three-dimensional flows around low-aspect-ratio flat-plate wings at
  low reynolds numbers.
\newblock {\em Journal of Fluid Mechanics}, 623:187--207, 2009.

\bibitem{eldredge2019leading}
Jeff~D Eldredge and Anya~R Jones.
\newblock Leading-edge vortices: mechanics and modeling.
\newblock {\em Annual Review of Fluid Mechanics}, 51:75--104, 2019.

\bibitem{kerstens2011closed}
Wesley Kerstens, Jens Pfeiffer, David Williams, Rudibert King, and Tim
  Colonius.
\newblock Closed-loop control of lift for longitudinal gust suppression at low
  reynolds numbers.
\newblock {\em AIAA journal}, 49(8):1721--1728, 2011.

\bibitem{perrotta2017unsteady}
Gino Perrotta and Anya~R Jones.
\newblock Unsteady forcing on a flat-plate wing in large transverse gusts.
\newblock {\em Experiments in Fluids}, 58(8):101, 2017.

\bibitem{fukami2020assessment}
Kai Fukami, Koji Fukagata, and Kunihiko Taira.
\newblock Assessment of supervised machine learning methods for fluid flows.
\newblock {\em Theoretical and Computational Fluid Dynamics}, pages 1--23,
  2020.

\bibitem{hou2019machine}
Wei Hou, Darwin Darakananda, and Jeff~D Eldredge.
\newblock Machine-learning-based detection of aerodynamic disturbances using
  surface pressure measurements.
\newblock {\em AIAA Journal}, 57(12):5079--5093, 2019.

\bibitem{kalman1960new}
Rudolph~Emil Kalman.
\newblock A new approach to linear filtering and prediction problems.
\newblock {\em Journal of Basic Engineering}, 1960.

\bibitem{asch2016data}
Mark Asch, Marc Bocquet, and Ma{\"e}lle Nodet.
\newblock {\em Data assimilation: methods, algorithms, and applications},
  volume~11.
\newblock SIAM, 2016.

\bibitem{evensen1994sequential}
Geir Evensen.
\newblock Sequential data assimilation with a nonlinear quasi-geostrophic model
  using monte carlo methods to forecast error statistics.
\newblock {\em Journal of Geophysical Research: Oceans}, 99(C5):10143--10162,
  1994.

\bibitem{da2018ensemble}
Andre~FC da~Silva and Tim Colonius.
\newblock Ensemble-based state estimator for aerodynamic flows.
\newblock {\em AIAA Journal}, 56(7):2568--2578, 2018.

\bibitem{da2020flow}
Andre~FC da~Silva and Tim Colonius.
\newblock Flow state estimation in the presence of discretization errors.
\newblock {\em Journal of Fluid Mechanics}, 890, 2020.

\bibitem{taira2007immersed}
Kunihiko Taira and Tim Colonius.
\newblock The immersed boundary method: a projection approach.
\newblock {\em Journal of Computational Physics}, 225(2):2118--2137, 2007.

\bibitem{taira2017modal}
Kunihiko Taira, Steven~L Brunton, Scott~TM Dawson, Clarence~W Rowley, Tim
  Colonius, Beverley~J McKeon, Oliver~T Schmidt, Stanislav Gordeyev, Vassilios
  Theofilis, and Lawrence~S Ukeiley.
\newblock Modal analysis of fluid flows: An overview.
\newblock {\em Aiaa Journal}, pages 4013--4041, 2017.

\bibitem{vonkarman38:1j}
T.~von K{\'a}rm{\'a}n and W.~R. Sears.
\newblock Airfoil theory for non-uniform motion.
\newblock {\em J. Aeronautical Sci.}, 5(10):379--390, 1938.

\bibitem{brown1955slender}
C.~E. Brown and W.~H. Michael.
\newblock Effect of leading edge separation on the lift of a delta wing.
\newblock {\em J. Aero. Sci.}, 21:690--694, 1954.

\bibitem{sarpkaya1975inviscid}
Turgut Sarpkaya.
\newblock An inviscid model of two-dimensional vortex shedding for transient
  and asymptotically steady separated flow over an inclined plate.
\newblock {\em Journal of Fluid Mechanics}, 68(1):109--128, 1975.

\bibitem{cottet2000vortex}
Georges-Henri Cottet and Petros~D. Koumoutsakos.
\newblock {\em Vortex methods: theory and practice}.
\newblock Cambridge university press, 2000.

\bibitem{eldredgebook}
Jeff~D. Eldredge.
\newblock {\em Mathematical Modeling of Unsteady Inviscid Flows}.
\newblock Springer International Publishing, 2019.

\bibitem{wang2013low}
Chengjie Wang and Jeff~D Eldredge.
\newblock Low-order phenomenological modeling of leading-edge vortex formation.
\newblock {\em Theoretical and Computational Fluid Dynamics}, 27(5):577--598,
  2013.

\bibitem{xiamohseni17}
X.~Xia and K.~Mohseni.
\newblock Unsteady aerodynamics and vortex-sheet formation of a two-dimensional
  airfoil.
\newblock {\em J. Fluid Mech.}, 830:439--478, 2017.

\bibitem{ramesh2014discrete}
Kiran Ramesh, Ashok Gopalarathnam, Kenneth Granlund, Michael~V Ol, and Jack~R
  Edwards.
\newblock Discrete-vortex method with novel shedding criterion for unsteady
  aerofoil flows with intermittent leading-edge vortex shedding.
\newblock {\em Journal of Fluid Mechanics}, 751:500--538, 2014.

\bibitem{darakananda2019versatile}
Darwin Darakananda and Jeff~D Eldredge.
\newblock A versatile taxonomy of low-dimensional vortex models for unsteady
  aerodynamics.
\newblock {\em Journal of Fluid Mechanics}, 858:917--948, 2019.

\bibitem{ramesh2018leading}
Kiran Ramesh, Kenneth Granlund, Michael~V Ol, Ashok Gopalarathnam, and Jack~R
  Edwards.
\newblock Leading-edge flow criticality as a governing factor in leading-edge
  vortex initiation in unsteady airfoil flows.
\newblock {\em Theoretical and Computational Fluid Dynamics}, 32(2):109--136,
  2018.

\bibitem{spalart1988vortex}
Philippe~R Spalart.
\newblock Vortex methods for separated flows.
\newblock Von Karman Institute for Fluid Dynamics, 1988.

\bibitem{sureshbabu2019model}
ArunVishnu SureshBabu, Kiran Ramesh, and Ashok Gopalarathnam.
\newblock Model reduction in discrete-vortex methods for unsteady airfoil
  flows.
\newblock {\em AIAA Journal}, 57(4):1409--1422, 2019.

\bibitem{darakananda2018data}
Darwin Darakananda, Andr{\'e} Fernando de~Castro da~Silva, Tim Colonius, and
  Jeff~D Eldredge.
\newblock Data-assimilated low-order vortex modeling of separated flows.
\newblock {\em Physical Review Fluids}, 3(12):124701, 2018.

\bibitem{bishop2001adaptive}
Craig~H Bishop, Brian~J Etherton, and Sharanya~J Majumdar.
\newblock Adaptive sampling with the ensemble transform kalman filter. part i:
  Theoretical aspects.
\newblock {\em Monthly weather review}, 129(3):420--436, 2001.

\bibitem{darakananda2017vortex}
Darwin Darakananda.
\newblock {\em Vortex Models for Data Assimilation}.
\newblock PhD thesis, UCLA, 2017.

\bibitem{burgers1998analysis}
Gerrit Burgers, Peter Jan~van Leeuwen, and Geir Evensen.
\newblock Analysis scheme in the ensemble kalman filter.
\newblock {\em Monthly weather review}, 126(6):1719--1724, 1998.

\bibitem{raanes2015improvements}
Patrick~N Raanes.
\newblock {\em Improvements to ensemble methods for data assimilation in the
  geosciences}.
\newblock PhD thesis, University of Oxford, 2015.

\bibitem{livings2008unbiased}
David~M Livings, Sarah~L Dance, and Nancy~K Nichols.
\newblock Unbiased ensemble square root filters.
\newblock {\em Physica D: Nonlinear Phenomena}, 237(8):1021--1028, 2008.

\bibitem{evensen2009ensemble}
Geir Evensen.
\newblock The ensemble kalman filter for combined state and parameter
  estimation.
\newblock {\em IEEE Control Systems Magazine}, 29(3):83--104, 2009.

\bibitem{sakov2008deterministic}
Pavel Sakov and Peter~R Oke.
\newblock A deterministic formulation of the ensemble kalman filter: an
  alternative to ensemble square root filters.
\newblock {\em Tellus A: Dynamic Meteorology and Oceanography}, 60(2):361--371,
  2008.

\bibitem{nerger2012unification}
Lars Nerger, Tijana Janji{\'c}, Jens Schr{\"o}ter, and Wolfgang Hiller.
\newblock A unification of ensemble square root kalman filters.
\newblock {\em Monthly Weather Review}, 140(7):2335--2345, 2012.

\bibitem{todter2015second}
Julian T{\"o}dter and Bodo Ahrens.
\newblock A second-order exact ensemble square root filter for nonlinear data
  assimilation.
\newblock {\em Monthly Weather Review}, 143(4):1347--1367, 2015.

\bibitem{whitaker2012evaluating}
Jeffrey~S Whitaker and Thomas~M Hamill.
\newblock Evaluating methods to account for system errors in ensemble data
  assimilation.
\newblock {\em Monthly Weather Review}, 140(9):3078--3089, 2012.

\bibitem{anderson1999monte}
Jeffrey~L Anderson and Stephen~L Anderson.
\newblock A monte carlo implementation of the nonlinear filtering problem to
  produce ensemble assimilations and forecasts.
\newblock {\em Monthly Weather Review}, 127(12):2741--2758, 1999.

\bibitem{pavliotis2014stochastic}
Grigorios~A Pavliotis.
\newblock {\em {Stochastic processes and applications: diffusion processes, the
  Fokker-Planck and Langevin equations}}.
\newblock Springer, 2014.

\bibitem{leonard:3j}
A.~Leonard.
\newblock Vortex methods for flow simulation.
\newblock {\em J.\ Comput.\ Phys.}, 37(3):289--335, 1980.

\bibitem{chorin1973numerical}
Alexandre~Joel Chorin.
\newblock Numerical study of slightly viscous flow.
\newblock {\em Journal of fluid mechanics}, 57(4):785--796, 1973.

\bibitem{long1988convergence}
Ding-Gwo Long.
\newblock Convergence of the random vortex method in two dimensions.
\newblock {\em Journal of the American Mathematical Society}, 1(4):779--804,
  1988.

\bibitem{Evensen2003TheImplementation}
Geir Evensen.
\newblock {The Ensemble Kalman Filter: Theoretical formulation and practical
  implementation}.
\newblock {\em Ocean Dynamics}, 53(4):343--367, 2003.

\bibitem{Law2015}
Kody Law, Andrew Stuart, and Konstantinos Zygalakis.
\newblock {\em Discrete Time: Filtering Algorithms}, pages 79--114.
\newblock Springer International Publishing, Cham, 2015.

\bibitem{liska2017fast}
Sebastian Liska and Tim Colonius.
\newblock A fast immersed boundary method for external incompressible viscous
  flows using lattice green's functions.
\newblock {\em Journal of Computational Physics}, 331:257--279, 2017.

\bibitem{ramesh2013unsteady}
Kiran Ramesh, Ashok Gopalarathnam, Jack~R Edwards, Michael~V Ol, and Kenneth
  Granlund.
\newblock An unsteady airfoil theory applied to pitching motions validated
  against experiment and computation.
\newblock {\em Theoretical and Computational Fluid Dynamics}, 27(6):843--864,
  2013.

\bibitem{houtekamer1998data}
Peter~L Houtekamer and Herschel~L Mitchell.
\newblock Data assimilation using an ensemble kalman filter technique.
\newblock {\em Monthly Weather Review}, 126(3):796--811, 1998.

\end{thebibliography}
